%% file: AdjointDMN.tex
\documentclass[a4paper,10pt]{article}

\usepackage[utf8]{inputenc}
\usepackage{authblk}

\usepackage{ifthen,ifpdf}

\ifpdf
\usepackage{hyperref}
\hypersetup{colorlinks=false,allcolors=blue}
\usepackage{hypcap}
\pdfpagewidth=\paperwidth
\pdfpageheight=\paperheight
\fi

\usepackage[utf8]{inputenc}
\usepackage[T1]{fontenc}

\usepackage{amsmath, amsthm, amssymb}
\usepackage{mathtools}
\usepackage{enumerate}
\usepackage{afterpage}
\usepackage{tabularx}
\usepackage{dsfont}

\usepackage{graphicx}

\usepackage{algorithm}
\usepackage{algpseudocode}
\usepackage{caption}
\usepackage[labelformat=simple]{subcaption}

\usepackage{color}
\usepackage{units}
\usepackage{array, multirow}
\usepackage{booktabs}

\newcolumntype{M}[1]{>{\vspace{3pt}\raggedleft\arraybackslash}m{#1}}

\usepackage{siunitx}[=v2]
\usepackage{tikz}
\usetikzlibrary{calc}%to move common legend between two graphs
\usetikzlibrary{matrix}
\usepackage[export]{adjustbox}
\usepackage{pgfplots}
\usepackage{pgfplotstable}
\usepgfplotslibrary{units}
\pgfplotsset{compat=1.9,samples=1000,every axis/.append style={
		font=\large,
		line width=1pt,
		tick style={line width=0.8pt}}}

\usetikzlibrary{arrows}
\usetikzlibrary{hobby}
\usetikzlibrary{calc}
\usetikzlibrary{arrows}
\usetikzlibrary{shapes}
\usetikzlibrary{positioning}
\usetikzlibrary{intersections}
\usetikzlibrary{decorations.pathreplacing}
\usetikzlibrary{patterns}
\usetikzlibrary{decorations.pathmorphing}
\usetikzlibrary{decorations.markings}
\usetikzlibrary{3d}
\usetikzlibrary{matrix}
\usetikzlibrary{spy}

\usepgfplotslibrary{polar}
\usepgfplotslibrary{external}
\usepgfplotslibrary{fillbetween}
\usepgfplotslibrary{groupplots}
\usepgfplotslibrary{patchplots}

\usepackage{cite}

\usepackage{anysize} 
\marginsize{2.5cm}{2.5cm}{2cm}{2cm}

\usepackage{wrapfig}

\def\XXint#1#2#3{{\setbox0=\hbox{$#1{#2#3}{\int}$}
		\vcenter{\hbox{$#2#3$}}\kern-.5\wd0}}

\usepackage{epstopdf}
\usepackage{physics}

\usepackage{printlen}

\usepackage{layouts}
\newcommand\inputpgf[2]{{
		\let\pgfimageWithoutPath\pgfimage
		\renewcommand{\pgfimage}[2][]{\pgfimageWithoutPath[##1]{#1/##2}}
		\input{#1/#2}
}}
\usepackage{import}
\usepgfplotslibrary{external}
\usepackage{longtable}
\usepackage{tikz-cd}
 
\newcommand{\DMN}[2]{#1\text{DMN}$_{#2}$}
\newcommand{\Diff}[2]{\frac{\mathrm{D} #1}{\mathrm{D} #2}}

\theoremstyle{plain}

\theoremstyle{remark}

\numberwithin{equation}{section}

\input{commands.tex}

\title{Learning Inelastic Homogenization with Memory: \\Deep Material Networks and Discrete Adjoints}

\author[1]{Binh Huy Nguyen}
\author[2,3,4,*]{Matti Schneider}
\affil[1]{Centre de mise en forme des mat\'eriaux (CEMEF), Mines Paris-PSL}
\affil[2]{University of Duisburg-Essen, Institute of Engineering Mathematics}
\affil[3]{Center for Nanointegration Duisburg-Essen (CENIDE)}
\affil[4]{Fraunhofer Institute for Industrial Mathematics ITWM, Kaiserslautern}
\affil[*]{correspondence to: \texttt{matti.schneider@uni-due.de}}

\date{\today}

\begin{document}
\maketitle 
%\printinunitsof{in}\prntlen{\textwidth}\prntlen{\textheight}
\begin{abstract}
\noindent Deep material networks (DMNs) provide mechanically structured surrogate models for computational homogenization. Traditionally, they are identified from elastic data and subsequently extrapolated to inelastic responses. This procedure leverages highly efficient linear solvers and provides robust extrapolation to different inelastic constitutive laws. However, selecting the appropriate linear elastic training data for a given inelastic constitutive behavior is non-trivial.\\
In this work, we identify the DMNs directly from macroscopic stress-strain trajectories. Applying automatic differentiation to such problems with history entails a huge memory demand, excluding most practically relevant problems. To mitigate, we derive a path-consistent discrete adjoint that avoids propagating parameter-wise state sensitivities by treating equilibrium and internal variable evolution over the complete loading history as coupled constraints.\\
For a short-fiber composite with an elastoplastic matrix, the adjoint approach reproduces the sensitivities obtained by direct differentiation and by an independent automatic-differentiation implementation, while it requires only one backward solve of the linearized state equations per time step and avoids the parameter-wise propagation of state sensitivities. Inelastic training also improves predictive accuracy: A rank-$5$ DMN trained on inelastic trajectories outperforms a deeper rank-$8$ DMN trained on elastic homogenization data for different loading paths. This advantage is retained under perturbations of the constituent inelastic material parameters. The resulting inelastic trained DMNs preserve the same formulation for online deployment and remain several orders of magnitude faster than the full-field homogenization solver.\\
 
	
\end{abstract}

%\t\tableofcontents
%\t\newpage

\newpage	
\section{Introduction}
\subsection{State of the art}
%\t\todo{ 
%\t	\begin{itemize}
%\t		\item FE2, ROM, E3C hyper-reduction
%\t		\item Vanilla DMN, direct DMN, IMN
%\t		\item Adjoint in optimization
%\t		\item Neural ODE
%\t	\end{itemize}		
%\t}
Computational homogenization provides physics-based constitutive models that account for the material microstructure without resolving the heterogeneities throughout the macroscopic structure, which would be prohibitively expensive~\cite{ZohdiWriggers2005}. It solves the microscopic equilibrium problem on a representative volume element (RVE) and assigns the resulting effective response to a material point of the macroscopic body~\cite{zhikov1994homogenization,cioranescu1999introduction,pavliotis2008multiscale}.
For linear elastic materials, the microscopic and macroscopic problems can be decoupled: The effective properties of a given RVE may be computed in advance, stored, and subsequently used in a homogeneous structural analysis. By contrast, the close relation between macroscopic inelastic behavior and the evolution of the microscopic constituents, for example, between macroscopic plastic strain and crystallographic slip~\cite{rice1970structure}, necessitates a coupled or nested two-scale formulation. In this concurrent setting, the microscopic internal variables evolve through the solution of a nonlinear cell problem at every macroscopic material point and load increment. Mature numerical techniques, most notably the finite element method (FEM)~\cite{guedes1990preprocessing,ZohdiWriggers2005,geers2010multi} and FFT-based solvers~\cite{moulinec1994fast,moulinec1998numerical,zeman2010accelerating}, see also Refs.~\cite{willot2015fourier,kabel2014efficient,schneider2017fft} and the review~\cite{schneider2021review}, are commonly employed as high-fidelity microscopic solvers in $\mathrm{FE}^2$~\cite{Smit1998,Feyel1999} and FE-FFT frameworks~\cite{kochmann2016two,kochmann2017efficient}.

Although cheaper than a fully resolved simulation, the repeated solution of nonlinear RVE problems remains computationally demanding, which limits concurrent multiscale simulations at the industrial scale and motivates reduced microscopic homogenizers that retain the fidelity of full-field models, particularly for inelastic materials. In the transformation field analysis (TFA) proposed by Dvorak~\cite{dvorak1992transformation}, the microstructure is partitioned into subdomains within which the inelastic strain is assumed to be uniform. Consequently, a fine partition may be required to achieve satisfactory accuracy~\cite{dvorak1992transformation,Fish1997}. To alleviate this limitation, Michel and Suquet~\cite{michel2003nonuniform} introduced the nonuniform transformation field analysis (NTFA), in which the spatiotemporal inelastic strain field is represented by spatially heterogeneous, time-independent modes and time-dependent activity coefficients. NTFA was applied to nonlinear composite structures~\cite{michel2004computational,fritzen2010three}, extended to linear viscoelastic composites~\cite{largenton2014extension}, and improved by mixed variational formulations~\cite{fritzen2013reduced} and series expansions~\cite{michel2016model}. In a similar spirit, Liu et al.~\cite{liu2016self,liu2018microstructural} proposed self-consistent clustering analysis, which approximates the Lippmann--Schwinger equation through $k$-means clustering combined with a self-consistent scheme in the spirit of Hill~\cite{hill1965self}. Schneider~\cite{schneider2019mathematical} provides a detailed account of the mathematical connections among these reduced-order homogenization methods. As noted by Michel and Suquet~\cite{michel2016model}, these approaches identify reduced bases for the internal variables, rather than for the displacement or velocity fields as in conventional projection-based model reduction. The latter category includes, for example, the proper-orthogonal-decomposition model of Yvonnet and He~\cite{yvonnet2007reduced}, as well as subsequent hyper-reduction techniques~\cite{hernandez2014high}, such as the empirical cubature method~\cite{hernandez2017dimensional,lange2024monolithic} and the empirically corrected cluster cubature~\cite{wulfinghoff2025homogenization,wulfinghoff2025e3c,wulfinghoff2026computational}, Hashin--Shtrikman-type reduced finite element formulations~\cite{wulfinghoff2018model}, and laminate-based reduced models on regular grids~\cite{kabel2016model}.

Neural networks emerged as versatile surrogates for inelastic material behavior and were already used in the 1990s to map strain sequences to stress sequences~\cite{ghaboussi1991knowledge,ghaboussi1999neural,furukawa1998implicit}. More recently, recurrent neural networks (RNNs) and variants designed specifically for sequential data, such as long short-term memory networks, were employed to represent path-dependent material responses~\cite{li2019machine,gorji2020potential,mozaffar2019deep}, where self-consistency~\cite{bonatti2022importance} and thermodynamic structure~\cite{masi2021thermodynamics} were identified as important ingredients. Neural ordinary differential equations with internal state variables~\cite{Jones_2022} and history-aware neural operators~\cite{guo2025history} pursue related goals, building upon general operator-learning architectures~\cite{lu2021learning,kovachki2023neural}. In computational homogenization, neural networks constructed from high-dimensional model representations were also used to replace full-field microscopic solvers at small~\cite{le2015computational} and finite strains~\cite{yvonnet2009numerically,nguyen2019surrogate}, to accelerate reduced-order models of woven composites~\cite{wu2021micro}, and Fourier neural operators were identified as universal linear elastic homogenizers~\cite{nguyen2026universal}.

Neural-network architectures are commonly adapted to the structure of the problem at hand. In image processing, for example, convolutional neural networks use local kernels that reflect correlations among neighboring pixels~\cite{lecun1998gradient}. Similarly, prior knowledge or inductive biases underlie the design of rotation- and translation-invariant networks~\cite{Bruna2013} and equivariant transformers~\cite{tai2019equivariant}. The same principle motivated physics-informed and physics-constrained architectures for micromechanical homogenization. Liu et al.~\cite{liu2019deep,liu2019exploring} introduced the deep material network (DMN), which approximates the heterogeneous RVE response through a hierarchy of laminates which serve as mechanistic building blocks. For each block, inelastic homogenization problems may be solved semi-analytically and thus cheaply, and the hierarchy of blocks progressively produces the effective response. Supervised training identifies the parameters - the volume fractions and rotated frames of each laminate - from constituent properties and effective stiffness or compliance tensors obtained by linear homogenization. The resulting network may then be applied to nonlinear and inelastic material behavior of the phases. Such rotated DMNs were extended to transfer learning~\cite{liu2019transfer}, to thermomechanical~\cite{shin2024deep,shin2024deepB} and damage problems~\cite{liu2020deep,liu2021cell}, and to uncertainty quantification~\cite{robertson2025microstructure}, while alternative network constructions were proposed to improve their accuracy~\cite{huang2022microstructure,shin2023deep,li2024micromechanics}, for instance graph-enhanced~\cite{jean2025graph} and micropolar~\cite{francis2025micropolar} variants. Rotated DMNs were also deployed for short fiber-reinforced composites in industrial settings~\cite{liu2020intelligent,wei2023ls}. Comprehensive overviews and comparisons are provided in Refs.~\cite{wan2024decoding,wei2026deep,he2026systematic}.

Gajek et al.~\cite{gajek2020micromechanics} provided both theoretical insights and practical improvements. They showed that every DMN inherits thermodynamic consistency from its phase materials, independently of the training quality, which provides a physical guarantee for every trained network. Moreover, they showed that the effective response of a material with vanishing memory is determined to first order by the linear elastic homogenization function. On the practical side, they introduced direct DMNs, which are parameterized by volume fractions and lamination directions alone. As a lamination direction lives on the two-dimensional unit sphere, whereas a general rotation requires three scalar parameters, direct DMNs have fewer training parameters than Liu-type DMNs. For the offline training, Gajek et al. used Milton's formula~\cite[Chap.~9]{Milton2022} for the laminate building blocks, ensuring a seamless integration into deep-learning frameworks. For the online phase, they introduced a flattened representation of the hierarchical laminate, which resembles a conventional discretization and leads to significantly higher speed-ups than hierarchical implementations. Direct DMNs were subsequently deployed within $\mathrm{FE}^2$ frameworks for elastoplastic and thermoplastic problems~\cite{gajek2021fe,gajek2022fe,gajek2021efficient}, and were also used for the inverse characterization of viscoplastic and creep behavior~\cite{dey2023rapid}. Further applications include highly nonlinear load cases of short fiber-reinforced thermoplastics~\cite{dey2024effectiveness}, the quantification of process-induced uncertainties~\cite{meyer2022probabilistic}, and fiber suspensions with infinite material contrast~\cite{sterr2025deep}, while the non-degeneracy of the underlying laminate building blocks was studied in Ref.~\cite{lendvai2026non}. More recently, Srinivas et al.~\cite{srinivas2026rapid} introduced a displacement-based formulation to accelerate offline training.
To further generalize material-network architectures, Nguyen and Noels introduced the interaction material network (IMN)~\cite{nguyen2022micromechanics,nguyen2022interaction}. Rather than relying exclusively on laminates as building blocks, the IMN employs a more general interaction mechanism, providing additional flexibility. Wei et al.~\cite{wei2025orientation} extended the IMN to polycrystalline materials by accounting for the crystal orientations.

\subsection{Contributions}
%\t\todo{
%\t\begin{itemize}
%\t	\item Efficient training technique
%\t	\item Better accuracy for online extrapolation
%\t	\item Trade-off between sampling and accuracy
%\t\end{itemize}
%\t}
Most previous studies identify DMNs from low-cost linear homogenization data and subsequently use them with nonlinear constituent laws, and this transfer to inelastic responses was demonstrated repeatedly. The central question is therefore how strongly the elastic objective constrains the history-dependent homogenization operator. Dey et al.~\cite{dey2022training} provide a particularly direct illustration: Networks with very similar elastic validation errors can show markedly different creep errors. Their inelastically informed strategy selects among stored elastic training iterates by an inelastic validation error, without forming an inelastic training gradient. Gajek et al.~\cite{gajek2023material} subsequently showed that elastic-to-viscoelastic transfer can be improved substantially by material-informed sampling and loss construction based on the constitutive structure of a Maxwell material. Shin et al.~\cite{shin2026unet} enriched the elastic training by local field information of the full-field simulations. These results demonstrate both the potential and the limitation of elastic identification: Transfer can be improved if additional constitutive information is available, but the resulting strategy is tied to the specific material class.

Direct identification from nonlinear trajectories was introduced for interaction-based material networks by Nguyen and Noels~\cite{nguyen2022interaction}. Their training setup is close to the one considered here: The loss is accumulated over homogenized response trajectories generated from monotonic proportional loading paths, and optional homogenized descriptors such as porosity can be included in the objective. The fitting-parameter gradient is obtained by direct implicit differentiation of the converged network equilibrium. For their monotonic loading paths, however, sensitivities of the previously converged internal variables are not propagated between increments. The intermediate states are treated as substepping points along the monotonic path. The resulting sensitivity is therefore local to the current converged step with respect to the inherited constitutive state.

The present work removes this restriction by treating the equilibrium equations and constitutive-update equations over the complete loading trajectory as coupled constraints. The dependence of the committed material state on the trainable parameters is retained explicitly from one increment to the next. In direct form, this approach produces a forward recursion for the complete state-sensitivity matrix. In adjoint form, the same history dependence is propagated backward through the subsequent-step coupling. This distinction is critical for general history-dependent loading, including unloading, cyclic, and non-proportional paths, where the current response cannot be differentiated independently of the previously converged state, in general.

Adjoint sensitivity analysis is additionally advantageous when a scalar trajectory objective depends on many trainable parameters because it avoids propagating one state-sensitivity column per parameter. The adjoint method is widely used in sensitivity analysis and PDE-constrained optimization~\cite{michaleris1994tangent,schwedes2017application,bui2026adjoints}. In scientific machine learning, it also provides the fundamental differentiation mechanism for neural ordinary differential equations~\cite{chen2018neural} and deep-equilibrium models~\cite{bai2019deep}. Inelastic trajectory data were used for training material networks before~\cite{nguyen2022interaction}. The novelty of the present work therefore lies in the path-consistent sensitivity formulation and its discrete-adjoint realization for DMN identification.

This work makes two principal contributions. First, it formulates inelastic DMN training as a history-dependent constrained optimization problem and derives both its complete direct sensitivity recursion and the corresponding discrete adjoint. The formulation requires only macroscopic stress-strain trajectories: The local internal variables remain governed by the constitutive evolution equations and need not be included as supervised labels. Second, it systematically compares elastic and inelastic training across multiple network ranks and training-set sizes, thereby quantifying their effects on predictive accuracy, generalization to unseen loading histories, offline training cost, and online acceleration. For the cross-rank comparison, we distinguish between in-distribution accuracy and out-of-distribution generalization.

The remainder of this article is organized as follows. Section~\ref{sec:DMN-theory} recalls the DMN representation and elastic training and develops the inelastic training together with its direct and adjoint sensitivities, including their numerical verification. Section~\ref{sec:examples} compares elastic and inelastic training for a short-fiber composite in terms of accuracy, computational cost, generalization to unseen loading paths and transfer to perturbed constituent properties. Section~\ref{sec:conclusions} summarizes the lessons learned. 

\section{Deep material networks and training strategies}
\label{sec:DMN-theory}
\subsection{DMN representation}
\label{sec:DMN-representation}
We begin by recalling the nonlinear homogenization problem approximated by the DMN, see Suquet~\cite{suquet1985elements} for background. We consider a microstructure $Y$ characterized by a heterogeneous generalized standard material (GSM)~\cite{halphen1975materiaux}, encoded by the free-energy density 
\begin{equation}\label{eq:setup-energy}
	\Psi: Y \times \Sym{d} \times Z \rightarrow \R_{\ge 0},\qquad\qty(\fy,\feps,\vb{z})\mapsto \Psi(\fy,\feps,\vb{z}),
\end{equation}
and the dissipation potential 
\begin{equation}\label{eq:setup-dissipation}
	\Phi: Y \times Z \rightarrow \R_{\ge 0},\qquad \qty(\fy,\dot{\vb{z}}) \mapsto \Phi(\fy,\dot{\vb{z}}),
\end{equation}	
where the symmetric second-order tensor $\feps\in\Sym{d}$ denotes the strain, while the quantities $\vb{z}$ and $\dot{\vb{z}}$ stand for an internal variable and its rate in a Banach space $Z$, respectively. Given a macroscopic strain path over the time interval $\qty[0,T]$,
\begin{equation}
	\label{eq:setup-strain-path}
	\bar{\feps}: [0,T] \rightarrow \Sym{d},
\end{equation}
and an initial state
 	\begin{equation}
 		\label{eq:setup-initial-state}
 		\vb{z}_0: Y \rightarrow Z,
 	\end{equation}
homogenization determines the effective stress as the volume average
\begin{equation}
	\label{eq:setup-effective-response}
	\bar{\fsigma}(t) := \mean{\pdv{\Psi}{\feps}\qty(\cdot,\bar{\feps}(t) + \nabla^s\fu(\cdot,t), \vb{z}(\cdot, t) )}_Y\qfor t \in \qty[0,T].
\end{equation}
Here, the displacement fluctuation
\begin{equation}
	\label{eq:setup-displacement}
	\fu: Y \times [0,T] \rightarrow \R^d
\end{equation}
and the internal variable $\vb{z}$ satisfy the balance of linear momentum and Biot's evolution equation,
 	\begin{subequations} 
 	\label{eq:setup-governing-equations}
 	\begin{align}
 		\label{eq:setup-balance-linear}
 		\div \qty[\pdv{\Psi}{\feps}\qty(\cdot, \bar{\feps} + \nabla^s\fu, \vb{z})] &= \vb{0}, \\
 		\label{eq:setup-Biot}
 		\pdv{\Psi}{\vb{z}}\qty(\fy,\bar{\feps}(t) + \nabla^s\fu(\fy,t), \vb{z}(\fy, t) ) + \pdv{\Phi}{\dot{\vb{z}}}\qty(\fy,\dot{\vb{z}}) &= 0 \qfor (\fy,t) \in Y \times \qty[0,T].
 	\end{align} 	
 	\end{subequations}
Equivalently, the unit cell defines a nonlinear, history-dependent homogenization operator $\mathcal{M}_Y$ that maps the macroscopic strain history up to the current time $t$, together with the initial microscopic state, to the current effective stress, see, for instance References~\cite{gajek2020micromechanics,gajek2022fe}, as
\begin{equation}
	\label{eq:setup-nonlinear-homogenization-function}
	\bar{\fsigma}(t) =\mathcal{M}_Y\qty(\left.\bar{\feps}\right|_{[0,t]},\vb{z}_0) \qfor t\in\qty[0,T].
\end{equation}

\begin{figure}[!h]
	\centering  
	\includegraphics[trim = 0 0 0 0, clip,width=0.6\textwidth]{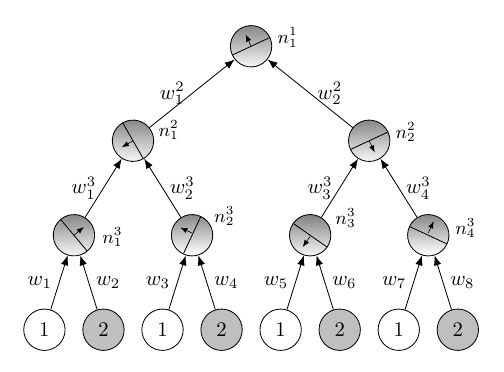} 
	\caption{Direct DMN of rank $3$ for a binary composite. The $M=2^3$ leaf nodes represent material points assigned to phase $1$ or phase $2$ and carry the corresponding constitutive response, including any evolving internal variables. The $N=M-1$ internal nodes define the hierarchical kinematics through the lamination directions $\vb{n}$, while the edge labels indicate the associated weights. Thus, the network topology is independent of whether the constituent response is elastic or inelastic.}
	\label{fig:binaryTreeDepth3}	
\end{figure}

The DMN provides a finite-dimensional, mechanically admissible surrogate for this operator. In the direct DMN~\cite{gajek2020micromechanics,gajek2021fe,gajek2022fe}, the network is organized as a hierarchy of directed laminates. For the two-phase material considered here, a network of rank $K$ forms a full binary tree with $M=2^K$ material points and $N=M-1$ internal nodes, as illustrated in Fig.~\ref{fig:binaryTreeDepth3}. The leaf nodes are assigned the constitutive laws of the two phases and carry the corresponding material-point histories. In particular, they are not restricted to linear elasticity. The internal nodes define the kinematic interactions through their lamination directions, while the material-point weights determine the relative contribution of the leaves.

Rather than evaluating the hierarchy recursively, the direct DMN can be flattened into a form analogous to a conventional discretization~\cite{gajek2020micromechanics}. Throughout this work, symmetric second-order tensors $\ftau \in \Sym{3}$ are identified with vectors in the space $\R^6$ via Mandel's notation
\begin{equation}
	\label{eq:setup-Mandel}
	\ftau \mapsto \qty(\tau_{11},\tau_{22},\tau_{33},\sqrt{2}\,\tau_{12},\sqrt{2}\,\tau_{13},\sqrt{2}\,\tau_{23})^T,
\end{equation}
so that the Frobenius inner product of tensors coincides with the Euclidean inner product of the associated vectors. All strain and stress vectors in this work, including the prescribed macroscopic strains and the effective stresses, refer to this representation. The vector collecting the strains at all material points is written as
\begin{equation}
	\label{eq:setup-strain-approximation}
	\vec{\feps} = A\bar{\feps} + D(\fw,\fn)\fu,
\end{equation}	 
where the matrix $A$ replicates the prescribed macroscopic strain at the material points, the matrix $D=D(\fw,\fn)$ represents the compatible symmetrized-gradient operator, and the vector $\fu$ collects the displacement jumps associated with the internal nodes. Compatibility follows because each local fluctuation strain is expressed as a symmetrized rank-one product of a displacement jump and a lamination direction. A general construction of the matrix $D$ is provided by Gajek et al.~\cite{gajek2020micromechanics}. For clarity, its explicit form for a rank-$3$ network is
\begin{equation}
	\label{eq:setup-Dmatrix-example-rank3}
	D = \mqty[c^2_2 B(n^1_1) & c^3_2 B(n^2_1) 	& \vb{0}  		 	& c^4_2 B(n^3_1)	& \vb{0}	& \vb{0} 	& \vb{0}\\
	c^2_2 B(n^1_1) & c^3_2 B(n^2_1) 	& \vb{0}   			& -c^4_1 B(n^3_1)	& \vb{0}	& \vb{0}	& \vb{0}\\
	c^2_2 B(n^1_1) & -c^3_1 B(n^2_1) 	& \vb{0}  			& \vb{0}			& c^4_4 B(n^3_2) & \vb{0} & \vb{0}\\
	c^2_2 B(n^1_1) & -c^3_1 B(n^2_1) 	& \vb{0}   			& \vb{0}			& -c^4_3 B(n^3_2) & \vb{0}	& \vb{0}\\
	-c^2_1 B(n^1_1) & \vb{0} 			& c^3_4 B(n^2_2)   	& \vb{0}			& \vb{0}		& c^4_6 B(n^3_3) & \vb{0}\\
	-c^2_1 B(n^1_1) & \vb{0} 			& c^3_4 B(n^2_2)    & \vb{0}			& \vb{0}		& -c^4_5 B(n^3_3) & \vb{0}\\
	-c^2_1 B(n^1_1) & \vb{0} 			& -c^3_3 B(n^2_2)   & \vb{0}			& \vb{0}	& \vb{0} & c^4_8 B(n^3_4)\\
	-c^2_1 B(n^1_1) & \vb{0} 			& -c^3_3 B(n^2_2)   & \vb{0}			& \vb{0} 	& \vb{0} & -c^4_7 B(n^3_4)\\
	],
\end{equation}
where the scalar $c_i^k$ denotes the volume fraction of the $i$-th node at level $k$ relative to its parent node,
\begin{equation}
	\label{eq:setup-volume-fraction}
	c^k_i := \frac{w^{k}_i}{w^k_{2j-1} + w^k_{2j}},\qquad j=\lceil i/2\rceil,
\end{equation}
and the matrix $B$ represents the symmetrized rank-one mapping
\begin{equation}
	\label{eq:setup-symmetrized-rank-one}
	\vb{u}\mapsto \frac{1}{2}\,\qty(\vb{n}\otimes\vb{u}+\vb{u}\otimes\vb{n}),
\end{equation}
and takes the form
\begin{equation}
	\label{eq:setup-B-matrix}
	B(\vb{n}) = \mqty[n_1	& 0 	& 0 	\\
	0	& n_2	& 0		\\
	0	& 0		& n_3		\\
	n_2/\sqrt 2	& n_1/\sqrt{2}	& 0	\\
	n_3/\sqrt 2	& 0	& n_1/\sqrt{2}		\\
	0	& n_3/\sqrt{2}	& n_2/\sqrt{2} ]
\end{equation}
in Mandel's notation~\eqref{eq:setup-Mandel}. At each material point, the local constitutive law maps the strain history and the material state to the stress. The total free energy of the DMN is the weighted sum of the material-point free energies. Stationarity with respect to the displacement jumps yields the discrete equilibrium equation
\begin{equation}
	\label{eq:setup-flattened-residual}
	D^T W \vec{\fsigma}\qty(A\bar{\feps} + D(\fw,\fn)\fu,z) = \vb{0},
\end{equation}
where the block-diagonal matrix $W=W(\fw)$ contains the material-point weights, the vector $z$ collects the material-point internal variables, and the vector $\vec{\fsigma}$ gathers the corresponding stresses. Note that Eq.~\eqref{eq:setup-flattened-residual}, introduced by Gajek et al.~\cite{gajek2020micromechanics}, provides a \textit{globally assembled}, rotation-free, FE-like formulation of the nonlinear online problem for DMNs, whereas the earlier DMNs~\cite{liu2019deep,liu2019exploring,liu2020deep} evaluated nonlinear response using recursive homogenization and de-homogenization along the hierarchical network. Ultimately, equations~\eqref{eq:setup-strain-approximation} and \eqref{eq:setup-flattened-residual} define the network representation used for both elastic and inelastic training. In particular, the matrices $D$ and $W$ describe the learned hierarchical kinematics and quadrature-like weights, whereas the constitutive laws assigned to the material points determine whether the resulting operator is elastic or inelastic.

\subsection{Network parametrization and elastic training}
\label{sec:DMN-parametrization-elastic-training}
The material network representation is determined by the nonnegative material-point weights $\fw$ and the unit lamination directions $\fn$. For optimization, these constrained quantities are expressed in terms of unconstrained trainable coordinates. The material-point weights are parametrized by unconstrained logits $\fxi\in\R^{M}$ through the floored softmax used in the present implementation,
\begin{equation}
	\label{eq:setup-parametrization-weights}
	w_i = w_{\min} + \qty(1-Mw_{\min}) \frac{\exp(\xi_i)}{\displaystyle\sum_{j=1}^{M}\exp(\xi_j)}, \qquad i=1,\ldots,M, \qquad w_{\min}=10^{-9}.
\end{equation}
Consequently, the conditions
\begin{equation}
	\label{eq:setup-weight-constraints}
	w_i\geq w_{\min}, \qquad i=1,\ldots,M, \qquad \text{and} \qquad \sum_{i=1}^{M}w_i=1
\end{equation}
are satisfied by construction. For every internal node, the unit lamination direction is represented in spherical coordinates~\cite{gajek2020micromechanics} as
\begin{equation}
	\label{eq:setup-parametrization-normals}
	\fn^{i}_{j} = \mqty[\sin \alpha^i_j \cos \beta^i_j \\
	\sin \alpha^i_j \sin \beta^i_j \\
	\cos \alpha^i_j],
\end{equation}
where, owing to periodicity and the sign invariance of the lamination direction, the angles may be treated as unconstrained real coordinates with representative ranges
\begin{subequations}
\begin{align}
	\label{eq:setup-parametrization-spherical-samplings}
	\alpha^i_j & \in [0,\pi), \\
	\beta^i_j & \in [0,2\pi).
\end{align}
\end{subequations}
For compactness, we collect the unconstrained weight variables and all angular coordinates into the vector
\begin{equation}
	\label{eq:setup-trainable-parameters}
	\ftheta := \qty(\fxi,\boldsymbol{\alpha},\boldsymbol{\beta}) \in \R^{N_\theta}
\end{equation}
as the trainable parameters to be determined from elastic and inelastic training strategies in the following sections. For given parameters $\ftheta$, the weights $\fw$ and the normal directions $\fn$ determine the DMN representation via the operators $D(\ftheta)$ and $W(\ftheta)$.

Elastic training is one identification strategy for this same DMN representation. If the constituent laws are linear elastic, every directed-laminate node admits an analytical homogenization map,
\begin{equation}
	\label{eq:setup-homogenized-parent-node}
	\C^{i}_{j} = \mathcal{G}\qty(\C^{i+1}_{2j-1},\C^{i+1}_{2j},c^{i+1}_{2j-1},c^{i+1}_{2j},\fn^{i}_{j}),
\end{equation}
where the fourth-order tensors $\C^{i+1}_{2j-1}$ and $\C^{i+1}_{2j}$ denote the homogenized stiffnesses of the two child nodes and the scalars $c^{i+1}_{2j-1}$ and $c^{i+1}_{2j}$ their relative volume fractions~\eqref{eq:setup-volume-fraction}. For a directed laminate, the parent stiffness $\C^{i}_{j}$ can be evaluated from Milton's formula~\cite[Chap.~9]{Milton2022}
\begin{equation}
	\label{eq:setup-linear-homogenization-function}
	\qty[\qty( \C^{i}_{j} - \C^{0} )^{-1} + \ffGamma^0\qty(\fn^i_j) ]^{-1} = c^{i+1}_{2j-1} \qty[\qty( \C^{i+1}_{2j-1} - \C^{0} )^{-1} + \ffGamma^0\qty(\fn^i_j) ]^{-1} + c^{i+1}_{2j} \qty[\qty( \C^{i+1}_{2j} - \C^{0} )^{-1} + \ffGamma^0\qty(\fn^i_j) ]^{-1}.
\end{equation}
Here, the fourth-order tensor $\C^0$ denotes a reference stiffness and the fourth-order tensor $\ffGamma^0(\fn)$ the associated laminate operator, which acts on a symmetric tensor $\ftau$ via $\ffGamma^0(\fn):\ftau := \sym\qty(\fK^0(\fn)^{-1}(\ftau\fn)\otimes\fn)$ in terms of the acoustic tensor $\fK^0(\fn)\fa := (\C^0:(\fa\otimes\fn))\fn$~\cite[Chap.~9]{Milton2022}. The reference stiffness $\C^0$ needs to be chosen such that all differences $\C - \C^0$ appearing in Milton's formula~\eqref{eq:setup-linear-homogenization-function} are sign-definite, for instance as a multiple of the identity on symmetric tensors whose modulus exceeds the largest eigenvalue of all stiffnesses involved, see Kabel et al.~\cite{kabel2015use} for a discussion.
This recursion is the linear-elastic specialization of the network, rather than its general definition. It may be evaluated recursively, through the flattened displacement formulation, or through other equivalent laminate formulas. A direct implementation of Milton's formula requires repeated fourth-order tensor inversions. As an alternative, Srinivas et al.~\cite{srinivas2026rapid} introduced a displacement-based laminate formulation to accelerate offline training.

In conventional elastic training, the trainable parameters $\ftheta$ are identified from \textit{linear responses}, particularly the effective stiffness tensors. To evaluate the quality of fit, different loss functions may be used. In the work at hand, we minimize the root mean square of the relative Frobenius error of the effective stiffness tensor,
\begin{equation}
	\label{eq:elastic_training_loss_function_DMN}
	\mathcal{L}^{}\qty(\ftheta) := \sqrt{ \frac{1}{N_s} \sum_{s=1}^{N_s} \qty( \frac{\|\bar{\C}^{\text{pred}}_s(\ftheta) - \bar{\C}^{\text{reference}}_s \|_F} {\| \bar{\C}^{\text{reference}}_s \|_F} )^2 } \longrightarrow \min_{\ftheta},
\end{equation}
where the integer $N_s$ refers to the number of elastic samples and the fourth-order tensors $\bar{\C}^{\text{pred}}_s$ and $\bar{\C}^{\text{reference}}_s$ stand for the predicted and reference effective stiffness tensors, respectively. Throughout this work,
$\|\cdot\|_F$ denotes the Frobenius norm of a tensor, matrix, or complete trajectory array, whereas $\|\cdot\|_2$ denotes the Euclidean norm of a vector. Because the elastic problem contains no evolving material history, the gradient follows directly from the chain rule in the form
\begin{equation}
	\label{eq:elastic_training_loss_function_DMN_gradient}
	\pdv{\mathcal{L}}{\ftheta} = \sum_{s=1}^{N_s}\pdv{\mathcal{L}}{\bar{\C}^{\text{pred}}_s}\pdv{\bar{\C}^{\text{pred}}_s}{\ftheta},
\end{equation}
and automatic differentiation may be applied to the algebraic laminate or displacement-based evaluation. This simpler differentiation problem does not imply that elastic identification is inexpensive. Generating one reference effective stiffness tensor requires several independent full-field linear homogenization problems, and the cost of sampling a sufficiently broad material-property space was reported to be significant for industrial-scale microstructures~\cite{srinivas2026deep}. The data-generation timings reported in the numerical examples below provide a direct comparison for the present problem.

\subsection{Inelastic training}
\label{sec:inelastic-training}
\subsubsection{Training formulation}
\label{sec:inelastic-training-formulation}
The objective of inelastic training is to identify the DMN directly from inelastic response trajectories. Whereas linear homogenization produces a single effective stiffness tensor, the nonlinear homogenization operator in Eq.~\eqref{eq:setup-nonlinear-homogenization-function} maps a prescribed strain history to an effective-stress history. Therefore, the trainable network parameters are obtained by minimizing a loss function defined directly in terms of the effective-stress trajectories,
\begin{equation}
	\label{eq:inelastic_training_loss_function}
	\mathcal{L}\qty(\ftheta) := \frac{1}{N_s}\sum_{s=1}^{N_s}\qty(\frac{1}{N_t} \sum_{n=1}^{N_t} \frac{ \| \bar{\fsigma}^{\text{pred}}_{s,n}\qty(\ftheta) - \bar{\fsigma}^{\text{reference}}_{s,n} \|_2 }{\| \bar{\fsigma}^{\text{reference}}_{s,n} \|_2 } ) \longrightarrow \min_{\ftheta},
\end{equation}
where the integers $N_s$ and $N_t$ denote the numbers of loading trajectories and time steps, respectively, while the vectors $\bar{\fsigma}^{\text{pred}}_{s,n}$ and $\bar{\fsigma}^{\text{reference}}_{s,n}$ stand for the effective stresses at time step $n$ predicted by the network and obtained from the reference solution, respectively. The sum starts at the first load step, so that the reference stresses do not vanish, and the relative error weights all load steps equally, irrespective of the stress level. The non-differentiability of the Euclidean norm at a vanishing residual was irrelevant in practice. In contrast to the test error~\eqref{eq:elastic_training_e_s} introduced in Section~\ref{sec:examples-setting}, which measures the deviation over the complete trajectory, the step-wise normalization prevents the late load steps with large stresses from dominating the training. Formally, the gradient follows from the relation
\begin{equation}
	\label{eq:inelastic_training_loss_function_DMN_gradient}
	\pdv{\mathcal{L}}{\ftheta} = \sum_{s=1}^{N_s}\sum_{n=1}^{N_t}\pdv{\mathcal{L}}{\bar{\fsigma}^{\text{pred}}_{s,n}}\Diff{\bar{\fsigma}^{\text{pred}}_{s,n}}{\ftheta}.
\end{equation}
The second factor, however, contains the implicit dependence of the complete time-discrete solution on the network parameters. At time step $n$, the vector $\fu_n \in \R^{3N}$ of displacement jumps and the vector $\fv_n \in \R^{M n_v}$ of material-point variables depend on the converged solution at time step $n-1$. Here, the vector $\fv_n$ collects the stresses and internal variables at all material points, i.e., $n_v = 6 + n_z$ scalar quantities per material point for $n_z$ scalar internal variables, and the matrix $\fS_{\sigma}$ extracts the stress components. Following Ref.~\cite{michaleris1994tangent}, we introduce the time-discrete global and local residuals
\begin{equation}
	\label{eq:inelastic-training-global-residual}
	\fR_n\qty(\fv_n;\ftheta) := D^T(\ftheta) W(\ftheta)\fS_{\sigma}\fv_n
\end{equation}
and
\begin{equation}
	\label{eq:inelastic-training-local-residual}
	\fH_n\qty(\fu_n,\fv_n,\fu_{n-1},\fv_{n-1};\ftheta) := \fH\qty(\fv_n,\feps_n,\fv_{n-1},\feps_{n-1}) \quad \text{with} \quad \feps_n := A\bar{\feps}_n+D(\ftheta)\fu_n,
\end{equation}
which take values in the spaces $\R^{3N}$ and $\R^{M n_v}$, respectively, and correspond to the balance of linear momentum~\eqref{eq:setup-balance-linear} and Biot's evolution equation~\eqref{eq:setup-Biot}. Thus, the identification of the DMN parameters is the history-dependent constrained optimization problem
\begin{equation}
	\label{eq:inelastic-training-constrained-optimization}
	\underset{\ftheta}{\text{minimize}}\;\mathcal{L}(\ftheta) \qquad\text{subject to}\qquad \fR_n=\vb{0},\quad\fH_n=\vb{0},\quad n=1,\ldots,N_t.
\end{equation}
Even though the loss~\eqref{eq:inelastic_training_loss_function} is defined only in terms of the effective stress, its dependence on the network parameters is retained through the complete sequence of inherited constitutive states. The coupled constraints in the optimization problem~\eqref{eq:inelastic-training-constrained-optimization} therefore represent the full time-discrete chain of dependencies and provide the basis for both path-consistent direct and adjoint sensitivities.
\subsubsection{Implicit adjoint differentiation}
\label{sec:inelastic-training-adjoint}
Before deriving the adjoint equations, we summarize the underlying strategy. The adjoint method differentiates the converged incremental equations, rather than the Newton or return-mapping iterations used to obtain their solution. Compared with applying reverse-mode automatic differentiation to the unrolled forward solver, this strategy offers several advantages. First, the memory demand is determined by the converged states of the trajectory, independently of the number of Newton and return-mapping iterations. Second, the computed gradient does not depend on the particular nonlinear solver, its initial guess or its iteration history. Third, the backward sweep requires one solve of the transposed linearized state equations per time step, independently of the number of trainable parameters, in contrast to the direct sensitivities discussed in Section~\ref{sec:inelastic-training-direct}. On the downside, the converged forward trajectory needs to be stored, the derivatives of the constitutive update with respect to the previous state need to be provided in addition to the conventional consistent tangent, and the computed gradient is accurate only up to the tolerance of the forward solver.

For ease of exposition, we consider a single loading trajectory, so that the contribution of a batch is obtained by summing the resulting gradients over all trajectories. At each time step, we gather the unknowns in the state vector
\begin{equation}
	\label{eq:impl-adj-collect-variables}
	\vb{q}_n := (\fu_n,\fv_n) \in \R^{3N} \times \R^{M n_v}
\end{equation}
and combine the two residuals into the function
\begin{equation}
	\label{eq:impl-adj-constraints}
	\fF_n\qty(\vb{q}_n,\vb{q}_{n-1};\ftheta) := \begin{bmatrix} \fR_n\qty(\fv_n;\ftheta)\\ \fH_n\qty(\fu_n,\fv_n,\fu_{n-1},\fv_{n-1};\ftheta) \end{bmatrix},
\end{equation}
so that the constraints read $\fF_n = \vb{0}$ for $n=1,\ldots,N_t$. The initial state $\vb{q}_0$ is prescribed and therefore independent of the trainable parameters.

We define the loss contribution at time step $n$ as
\begin{equation}
	\label{eq:impl-adj-loss-at-n}
	L_n(\ftheta) := \frac{1}{N_t}\frac{ \| \bar{\fsigma}_{n}\qty(\ftheta) - \bar{\fsigma}^{\text{reference}}_{n} \|_2 }{\| \bar{\fsigma}^{\text{reference}}_{n} \|_2 },
\end{equation}
in terms of the effective stress
\begin{equation}
	\label{eq:impl-adj-stress-extraction}
	\bar{\fsigma}_n(\ftheta) := \bar{A}(\ftheta)\fS_{\sigma}\fv_n
\end{equation}
at time step $n$, computed with the help of the weighted averaging operator $\bar{A}(\ftheta)$. The accumulated loss over the trajectory defines the objective function
\begin{equation}
	\label{eq:impl-adj-traj-loss}
	\mathcal{J}(\ftheta) := \sum_{n=1}^{N_t}L_n(\ftheta).
\end{equation}
For the constrained optimization problem~\eqref{eq:inelastic-training-constrained-optimization}, we introduce the adjoint Lagrangian
\begin{equation}
	\label{eq:inelastic_training_adjoint_objective_function}
	\mathcal{L}_{\mathrm{adj}} := \mathcal{J}(\ftheta) -\sum_{n=1}^{N_t}\vb*{\lambda}_n^T\fF_n = \sum_{n=1}^{N_t}L_n -\sum_{n=1}^{N_t}\fgam_n^T\fR_n -\sum_{n=1}^{N_t}\feta_n^T\fH_n,
\end{equation}
where the vectors $\vb*{\lambda}_n=(\fgam_n,\feta_n) \in \R^{3N} \times \R^{M n_v}$ denote the adjoint variables associated with the global residual~\eqref{eq:inelastic-training-global-residual} and the local residual~\eqref{eq:inelastic-training-local-residual}, respectively. In particular, the adjoint variables live in the same spaces as the corresponding unknowns, i.e., the adjoint variable $\fgam_n$ is an adjoint displacement-jump vector, whereas the adjoint variable $\feta_n$ is associated with the stresses and internal variables at the material points. For the converged forward solution, the constraints $\fF_n = \vb{0}$ are satisfied, such that the objective function $\mathcal{J}$ and the adjoint Lagrangian $\mathcal{L}_{\mathrm{adj}}$ have the same sensitivity. Differentiating the adjoint Lagrangian~\eqref{eq:inelastic_training_adjoint_objective_function} with respect to the parameter vector $\ftheta$ and grouping the terms multiplying the state sensitivities leads to the relation
\begin{equation}
	\label{eq:inelastic_training_adjoint_DL}
	\begin{split}
	\Diff{\mathcal{J}}{\ftheta} ={}&\sum_{n=1}^{N_t}\qty( \pdv{L_n}{\ftheta} -\vb*{\lambda}_n^T\pdv{\fF_n}{\ftheta})\\
	&+\sum_{n=1}^{N_t}\qty[ \pdv{L_n}{\vb{q}_n} -\vb*{\lambda}_n^T\pdv{\fF_n}{\vb{q}_n} -\vb*{\lambda}_{n+1}^T\pdv{\fF_{n+1}}{\vb{q}_n} ]\Diff{\vb{q}_n}{\ftheta},
	\end{split}
\end{equation}
where the terminal adjoint variable is set to $\vb*{\lambda}_{N_t+1} := \vb{0}$. Here and below, partial derivatives with respect to the parameter vector $\ftheta$ are evaluated at the current time step $n$, with the state variables held fixed. The state-sensitivity terms in the relation~\eqref{eq:inelastic_training_adjoint_DL} vanish if the adjoint variables satisfy the equations
\begin{equation}
	\label{eq:inelastic_training_adjoint_DR_DH}
	\qty(\pdv{\fF_n}{\vb{q}_n})^T\vb*{\lambda}_n = \qty(\pdv{L_n}{\vb{q}_n})^T -\qty(\pdv{\fF_{n+1}}{\vb{q}_n})^T\vb*{\lambda}_{n+1}, \qquad n=N_t,\ldots,1.
\end{equation}
These adjoint equations are solved backward in time. The term involving the residual $\fF_{n+1}$ transfers sensitivity from the subsequent time step to the current one and thereby retains the complete dependence on the constitutive history.

In view of the definition~\eqref{eq:impl-adj-constraints}, we introduce the four blocks
\begin{subequations}
	\label{eq:inelastic_training_Jacobian_blocks}
	\begin{align}
		\fH^{\fv}_n&:=\pdv{\fH_n}{\fv_n}, \\
		\fH^{\feps}_n&:=\pdv{\fH_n}{\feps_n},\\
		\fH^{\fv,-}_n&:=\pdv{\fH_n}{\fv_{n-1}}, \\
		\fH^{\feps,-}_n&:=\pdv{\fH_n}{\feps_{n-1}}
	\end{align}
\end{subequations}
of the linearized constitutive residual. For a loss that depends on the state only through the effective stress, the adjoint equations~\eqref{eq:inelastic_training_adjoint_DR_DH} turn into the coupled global-local adjoint system
\begin{equation}
	\label{eq:inelastic_training_adjoint_coupled_equations}
	\left\{
	\begin{aligned}
	\fgam_n^T D^T W\fS_{\sigma} +\feta_n^T\fH^{\fv}_n +\feta_{n+1}^T\fH^{\fv,-}_{n+1} &=\pdv{L_n}{\bar{\fsigma}_n}\,\bar{A}\fS_{\sigma},\\
	\feta_n^T\fH^{\feps}_nD +\feta_{n+1}^T\fH^{\feps,-}_{n+1}D &=\vb{0},
	\end{aligned}
	\right. \qquad n=N_t,\ldots,1,
\end{equation}
with the convention $\feta_{N_t+1} := \vb{0}$ for the local adjoint variable. All network operators in the adjoint system~\eqref{eq:inelastic_training_adjoint_coupled_equations}, including the operators $D$, $W$ and $\bar{A}$, are evaluated at the current parameter vector $\ftheta$. The adjoint sweep is evaluated on the converged forward trajectory. At each time step, the constitutive Jacobian blocks are re-evaluated analytically at the stored converged state, and the adjoint system~\eqref{eq:inelastic_training_adjoint_coupled_equations} is solved backward from $n=N_t$ to $n=1$. The sweep requires one coupled adjoint solution per time step, and the number of these solutions is independent of the number of trainable network parameters. For the $J_2$ plasticity model used in the numerical examples, these constitutive Jacobian blocks are obtained analytically from the elastic predictor and plastic return-mapping relations~\cite{michaleris1994tangent}.
Substituting the adjoint variables obtained from the adjoint system~\eqref{eq:inelastic_training_adjoint_coupled_equations} into the relation~\eqref{eq:inelastic_training_adjoint_DL} eliminates the state sensitivities, leaving only explicit derivatives of the network operators. For an arbitrary parameter perturbation $\delta\ftheta \in \R^{N_\theta}$, the corresponding directional derivative reads
\begin{equation}
	\label{eq:inelastic_training_adjoint_DL_2}
	\begin{split}
	\mathrm{D}\mathcal{J}(\ftheta)[\delta\ftheta] =\sum_{n=1}^{N_t}\Big\{& \pdv{L_n}{\bar{\fsigma}_n} \qty[\delta\bar{A}\,\fS_{\sigma}\fv_n]\\
	&-\fgam_n^T\qty[ (\delta D)^T W\fS_{\sigma}\fv_n +D^T(\delta W)\fS_{\sigma}\fv_n]\\
	&-\feta_n^T\qty[ \fH^{\feps}_n(\delta D)\fu_n +\fH^{\feps,-}_n(\delta D)\fu_{n-1}] \Big\},
	\end{split}
\end{equation}
where the matrices $\delta D$, $\delta W$ and $\delta\bar{A}$ denote the directional derivatives of the operators $D$, $W$ and $\bar{A}$ at the parameter vector $\ftheta$ in the direction $\delta\ftheta$. Each term in the directional derivative~\eqref{eq:inelastic_training_adjoint_DL_2} is a vector--Jacobian product with respect to the parameter vector $\ftheta$, so that the loss gradient can be evaluated without forming the parameter-wise sensitivities $\mathrm{D}\fu_n/\mathrm{D}\ftheta$ and $\mathrm{D}\fv_n/\mathrm{D}\ftheta$. Automatic differentiation is used only for the explicit parameterizations of the operators $D(\ftheta)$, $W(\ftheta)$, and $\bar{A}(\ftheta)$, while the constitutive blocks are supplied by the material update.
%	The finite-difference verification below therefore includes a perturbation-size stability check, which also exposes sensitivity to branch crossings. In particular, the block $\fB_n$ describes the dependence on the current strain, while the blocks $\fC_n$ and $\fE_n$ propagate the dependence on the state from the preceding time step. Thus, the reverse sweep differentiates the converged constitutive equations at each time step rather than the iterations of the local solver. The cost of this sweep is governed by one coupled adjoint solution per time step and does not grow with the number of trainable network parameters.	
\subsubsection{Implicit direct differentiation}
\label{sec:inelastic-training-direct}
For reference and for verification of the implicit adjoint formulation, we also consider direct differentiation of the converged time-discrete problem. For a single trajectory, differentiating the objective function~\eqref{eq:impl-adj-traj-loss} with respect to the parameter vector $\ftheta$ yields the relation
\begin{equation}
	\label{eq:direct_loss_gradient}
	\Diff{\mathcal{J}}{\ftheta} = \sum_{n=1}^{N_t} \pdv{L_n}{\bar{\fsigma}_n} \Diff{\bar{\fsigma}_n}{\ftheta},
\end{equation}
where the direct sensitivity of the effective stress~\eqref{eq:impl-adj-stress-extraction} follows as
\begin{equation}
	\label{eq:direct_effective_stress_sensitivity}
	\Diff{\bar{\fsigma}_n}{\ftheta} = \Diff{\bar{A}}{\ftheta} \fS_{\sigma}\fv_n + \bar{A}\fS_{\sigma} \Diff{\fv_n}{\ftheta}.
\end{equation}
In view of the definition~\eqref{eq:impl-adj-collect-variables} of the state vector, we introduce the direct state-sensitivity matrix
\begin{equation}
	\label{eq:direct_state_sensitivity_matrix}
	\fQ_n := \Diff{\vb{q}_n}{\ftheta} \in \R^{(3N + M n_v) \times N_\theta}.
\end{equation}
Differentiating the converged constraints $\fF_n = \vb{0}$ with respect to the parameter vector $\ftheta$ leads to the equation
\begin{equation}
	\label{eq:direct_sensitivity_equation}
	\pdv{\fF_n}{\vb{q}_n}\fQ_n + \pdv{\fF_n}{\vb{q}_{n-1}}\fQ_{n-1} + \pdv{\fF_n}{\ftheta} =\vb{0},
\end{equation}
and therefore the recursion
\begin{equation}
	\label{eq:direct_sensitivity_recursion}
	\fQ_n=-\left(\pdv{\fF_n}{\vb{q}_n} \right)^{-1} \left[ \pdv{\fF_n}{\ftheta} + \pdv{\fF_n}{\vb{q}_{n-1}}\fQ_{n-1} \right].
\end{equation}
The recursion~\eqref{eq:direct_sensitivity_recursion} is evaluated forward in time and retains the complete dependence on the material history. In particular, the term $(\partial\fF_n/\partial\vb{q}_{n-1})\fQ_{n-1}$ propagates the sensitivity of the committed state from the preceding increment. Omitting this term yields a step-local sensitivity with respect to the inherited state, which is the simplification used by Nguyen and Noels~\cite{nguyen2022interaction} for their monotonic training paths. The direct formulation is mathematically straightforward, but the state-sensitivity matrix $\fQ_n$ contains one sensitivity column for every design variable and must be propagated through the complete loading history. Since only the matrix $\fQ_{n-1}$ enters the recursion~\eqref{eq:direct_sensitivity_recursion}, the memory demand of the direct approach is of the order $\mathcal{O}((3N + M n_v)\,N_\theta)$. More importantly, every time step requires the solution of a linear system with the matrix $\partial\fF_n/\partial\vb{q}_n$ for $N_\theta$ right-hand sides, so that the computational effort grows linearly with the number of trainable parameters. With the present parametrization, a rank-$K$ binary DMN has $N_{\theta}=M+2N$ trainable coordinates, i.e., $N_{\theta}=190$ at rank $6$ and $N_{\theta}=766$ at rank $8$. For comparison, the largest interaction-based material network reported by Nguyen and Noels contains $415$ fitting parameters~\cite{nguyen2022interaction}. In contrast, the adjoint system~\eqref{eq:inelastic_training_adjoint_coupled_equations} requires a single solution with the transposed matrix per time step, independently of the number $N_\theta$ of trainable parameters, at the expense of storing the converged forward trajectory, i.e., a memory demand of the order $\mathcal{O}(N_t\,(3N + M n_v))$. For the network sizes considered here, both memory demands are moderate, and the advantage of the adjoint formulation lies in the number of linear solves per gradient evaluation. The remaining parameter-dependent work consists of the vector--Jacobian products with the explicit parameterizations of the network operators, which are evaluated by reverse-mode automatic differentiation.

\subsubsection{Gradient verification}
\label{sec:inelastic-training-verification}
\begin{figure}[!h]
	\centering
	\includegraphics[height=0.03\textheight]{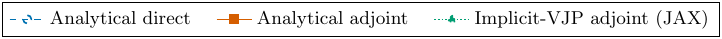}\\[2pt]
	\begin{subfigure}[t]{0.48\textwidth}
		\centering
		\includegraphics[height=5.4cm]{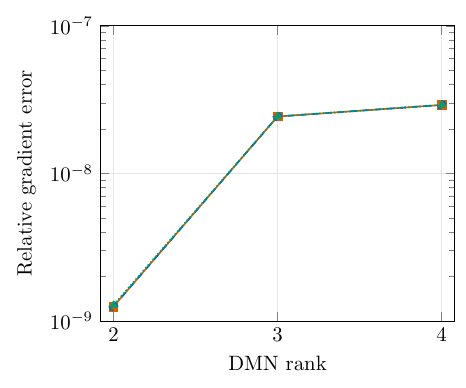}
		\caption{Complete trajectory gradient\\ for DMN ranks $2$--$4$}
		\label{fig:gradient_verification_rank}
	\end{subfigure}
	\hfill
	\begin{subfigure}[t]{0.48\textwidth}
		\centering
		\includegraphics[height=5.4cm]{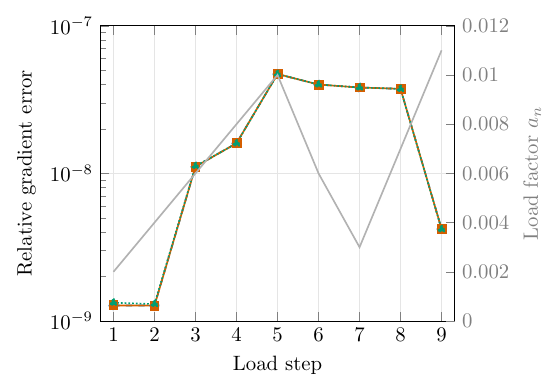}
		\caption{Gradient per load step for rank $4$,\\ load factor $a_n$ in gray}
		\label{fig:gradient_verification_step}
	\end{subfigure}
	\caption{Relative Euclidean error~\eqref{eq:gradient-ver-relative-error} of the gradients with respect to the centered finite-difference reference. The three implementations overlap.}
	\label{fig:gradient_verification}
\end{figure}

We verified three implementations against centered finite differences using a deterministic single-trajectory benchmark, independently of the training procedure: The direct differentiation of Section~\ref{sec:inelastic-training-direct}, the discrete adjoint of Section~\ref{sec:inelastic-training-adjoint}, and an independently implemented implicit reverse-mode realization based on JAX \texttt{custom\_vjp}~\cite{jax2018github}. The latter realization assigns a custom vector--Jacobian product to the converged nonlinear DMN step: Its reverse rule solves the transposed reduced tangent system and applies the residual pullback, without differentiating through the Newton iterations. It therefore realizes the same implicit discrete-adjoint relation as the adjoint equations~\eqref{eq:inelastic_training_adjoint_DR_DH}, but provides an independent automatic-differentiation implementation for verification. For this purpose, we prescribed the proportional macroscopic strain history
\begin{equation}
	\label{eq:gradient-ver-strain-path}
	\bar{\feps}_n := a_n\mqty[1 & -0.5 & -0.5 & 0.12 & 0 & 0]^T,
\end{equation}
in Mandel's notation~\eqref{eq:setup-Mandel}, where the load factors take the values
\begin{equation}
	\label{eq:gradient-ver-strain-magnitude}
	a_n\in\{0,\,0.002,\,0.004,\,0.006,\,0.008,\,0.010,\,0.006,\,0.003,\,0.007,\,0.011\}.
\end{equation}
The trajectory therefore contains loading, unloading, and reloading, such that the sensitivity calculation includes the propagation of the constitutive history. The same constituent material parameters as in the computational examples section were employed. For each network rank, a deterministic parameter initialization was used to compute a target effective-stress trajectory $\bar{\fsigma}^{\mathrm{target}}_n$. To obtain a nonzero objective and gradient without performing optimization, the reference trajectory was constructed as
\begin{equation}
	\label{eq:gradient-ver-reference-stress-trajectory}
	\bar{\fsigma}^{\mathrm{reference}}_n :=0.97\,\bar{\fsigma}^{\mathrm{target}}_n +10^{-5}\,\mqty[1 & -0.5 & 0.25 & 0.1 & -0.05 & 0.075]^T,\qq{for} n=1,\ldots,9,
\end{equation}
starting from the stress-free initial state at $n=0$. The complete trajectory gradient was approximated component-wise by centered finite differences of the trajectory objective in Eq.~\eqref{eq:impl-adj-traj-loss},
 	\begin{equation}
 		\label{eq:gradient-ver-finite-difference}
 		g_i^{\mathrm{FD}} := \frac{ \mathcal{J}(\ftheta+h_i\vb{e}_i)- \mathcal{J}(\ftheta-h_i\vb{e}_i)}{2h_i},
\end{equation}
where the vector $\vb{e}_i$ denotes the unit vector associated with the $i$-th trainable parameter $\theta_i$, and the scalar $h_i$ refers to the step size
\begin{equation}
	\label{eq:gradient-ver-step-size}
	h_i:=10^{-6}\,\max\qty(1,|\theta_i|).
\end{equation}
Collecting the components $g_i^{\mathrm{FD}}$ yields the finite-difference trajectory gradient $\vb{g}^{\mathrm{FD}}$. We also repeated the central finite-difference calculation with the relative perturbation $10^{-5}$ as a stability check of the finite-difference reference but do not show the result here. For each network rank, we consider the relative error
\begin{equation}
	\label{eq:gradient-ver-relative-error}
	e_{\mathrm{grad}} := \frac{\|\vb{g}-\vb{g}^{\mathrm{FD}}\|_2}{\|\vb{g}^{\mathrm{FD}}\|_2},
\end{equation}
where the vector
\begin{equation}
	\label{eq:verification-g}
	\vb{g}:=\Diff{\mathcal J}{\ftheta}
\end{equation}
denotes the complete trajectory gradient obtained by the direct formulation, the analytical adjoint, or the implicit-VJP adjoint. Figure~\ref{fig:gradient_verification_rank} presents the relative Euclidean error~\eqref{eq:gradient-ver-relative-error} of the complete parameter-gradient vector for a given network rank. For ranks $2$--$4$, all three implementations agree with the centered finite-difference reference to approximately $10^{-9}$--$10^{-8}$. The analytical and implicit-VJP adjoint curves are practically indistinguishable, as expected because they realize the same discrete-adjoint sensitivity relation through independent derivative implementations. Their agreement with the direct sensitivity and finite-difference references confirms the consistent treatment of the history dependence in the adjoint equations~\eqref{eq:inelastic_training_adjoint_DR_DH}. Figure~\ref{fig:gradient_verification_step} resolves the error of the gradient of the loss contribution $L_n$ for every load step $n$ at rank $4$. The error stays between $10^{-9}$ and $5 \times 10^{-8}$ along the complete trajectory. It increases during the initial loading up to step $5$, but it does not grow further during unloading in the steps $6$ and $7$ and reloading in the steps $8$ and $9$, and all three implementations agree at every load step. Hence, the sensitivity of the inherited constitutive state is propagated consistently from one increment to the next.
 
We restrict the numerical comparison to the accuracy of the gradients. The computational complexity of the direct and the adjoint approaches is discussed in Section~\ref{sec:inelastic-training-direct}.
 
\section{Computational examples}
\label{sec:examples}
\subsection{Problem setting}
\label{sec:examples-setting}

The numerical study compares elastic and inelastic training for a short-fiber composite. The microstructure, reference data, training configurations, and error measures are specified below.
\begin{figure}[!h]
	\centering
	\begin{minipage}[t]{0.31\textwidth}
		\centering
		\parbox[b][4.3cm][c]{\linewidth}{\centering \includegraphics[trim = 130 50 1280 420, clip,width=\linewidth,height=4.2cm,keepaspectratio]{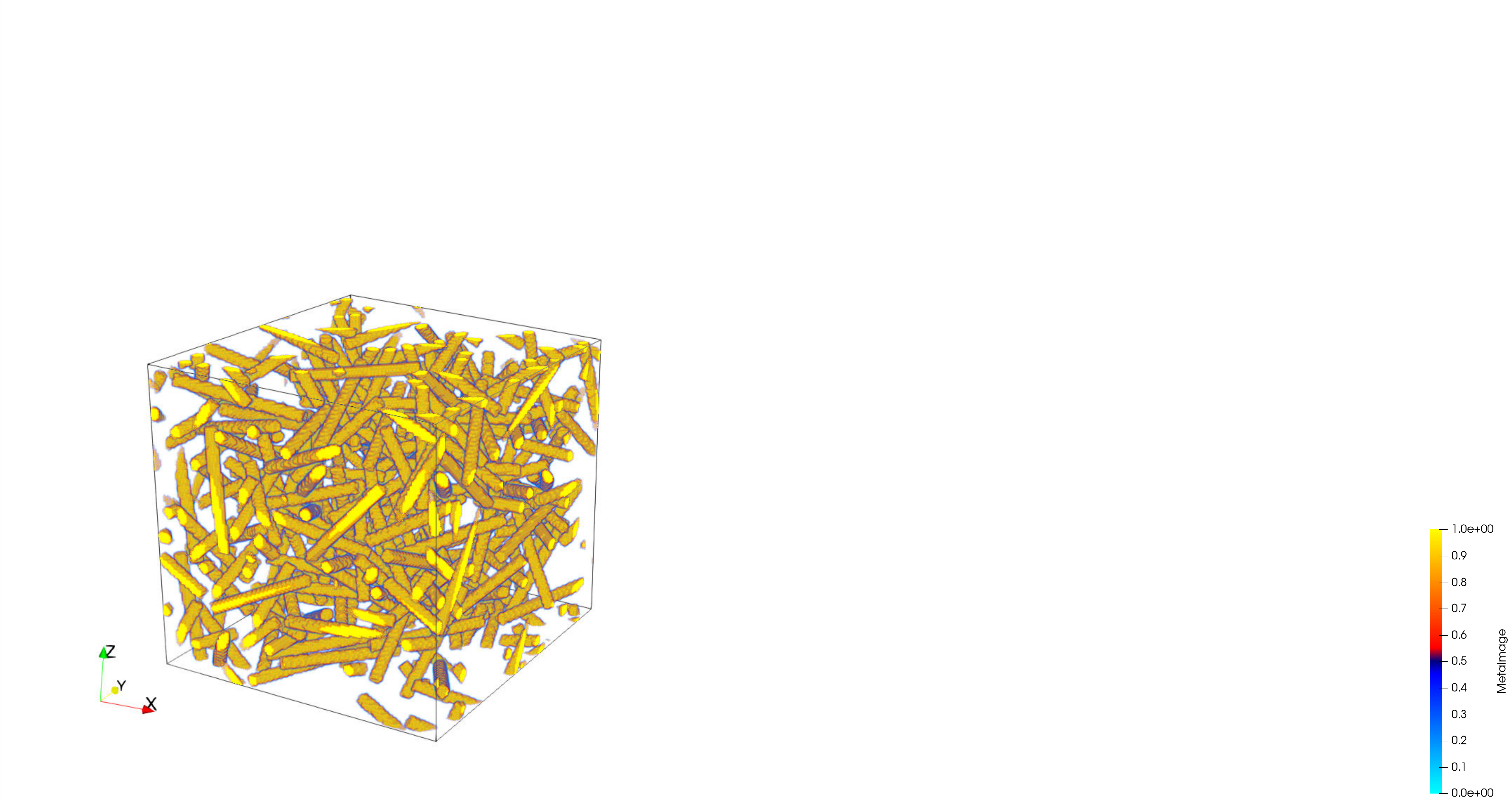}}
		\subcaption{Voxelized short-fiber composite\\ with $128^3$ voxels}
		\label{fig:problem_setup_fiber128}
	\end{minipage}
	\hfill
	\begin{minipage}[t]{0.31\textwidth}
		\centering
		\parbox[b][4.3cm][c]{\linewidth}{\centering \includegraphics[width=\linewidth,height=4.2cm,keepaspectratio]{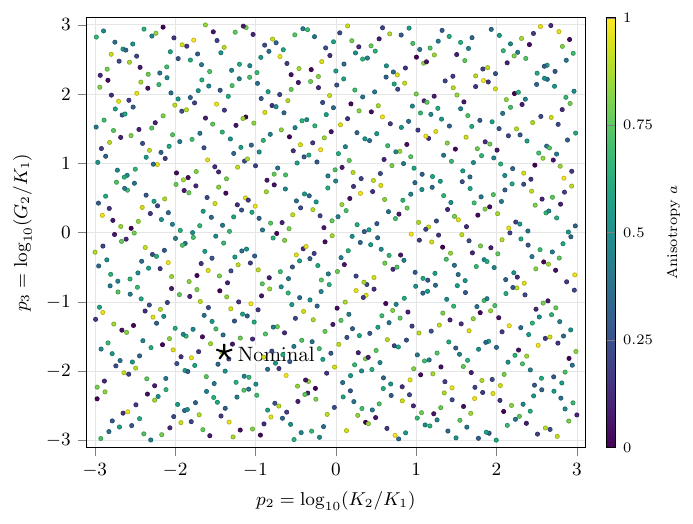}}
		\subcaption{Elastic data: stiffness contrasts}
		\label{fig:problem_setup_elastic_contrast}
	\end{minipage}
	\hfill
	\begin{minipage}[t]{0.31\textwidth}
		\centering
		\parbox[b][4.3cm][c]{\linewidth}{\centering \includegraphics[width=\linewidth,height=4.2cm,keepaspectratio]{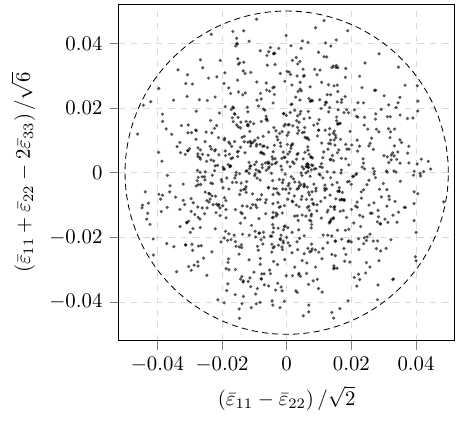}}
		\subcaption{Inelastic data: projected terminal strains, $R^{\max}=0.05$ (dashed)}
		\label{fig:problem_setup_inelastic_sampling}
	\end{minipage}
		\caption{Problem setting: microstructure and sampled training data.}
		\label{fig:problem_setup}		
	\end{figure}

\paragraph{Microstructure and reference solver.}
We consider a two-phase composite representative of a polymer matrix reinforced by short E-glass fibers of length $100\,\mu\mathrm{m}$ and diameter of $10\,\mu\mathrm{m}$, dispersed in an isotropic way encoded via the corresponding fiber-orientation tensor~\cite{montgomery2011fast}, with a fiber volume fraction of $10\,\%$. The microstructure, shown in Fig.~\ref{fig:problem_setup_fiber128}, was generated by the sequential addition and migration algorithm~\cite{schneider2017sequential}, with an edge length of $256\,\mu\mathrm{m}$ and was voxelized with a resolution of $2\,\mu\mathrm{m}$, resulting in a $128^3$ voxel grid. Subsequently, the resulting microstructure was discretized on a rotated staggered grid~\cite{willot2015fourier} and solved by an in-house FFT-based homogenization solver~\cite{schneider2019barzilai}. For the elastic data set, the problem was solved by the linear conjugate gradient method~\cite{zeman2010accelerating}, whereas the Eyre--Milton solver~\cite{eyre1999fast,schneider2019polarization} with tolerance of $10^{-5}$ was selected for the inelastic case.  

\paragraph{Elastic data.}
For elastic identification, we use $1000$ pairs of constituent stiffness tensors following the orthotropic sampling of Gajek et al.~\cite{gajek2021fe}. In Mandel notation, the two constituent tensors are
\begin{equation}
    \mathbf{C}_1=3K_1\mathbf{P}_{\mathrm{vol}}+2G_1\mathbf{P}_{\mathrm{dev}},
    \qquad
    \mathbf{C}_2=3K_2\mathbf{P}_{\mathrm{vol}}+
    2G_2\bigl(\mathbf{P}_{\mathrm{dev}}-a\,\mathbf{N}'\otimes\mathbf{N}'\bigr),
\end{equation}
where $\mathbf{P}_{\mathrm{vol}}$ and $\mathbf{P}_{\mathrm{dev}}$ are the volumetric and deviatoric projectors, respectively, and $\mathbf{N}'$ is a rotated, unit-norm deviatoric tensor. We set $K_1=1\,\mathrm{GPa}$, $G_1=10^{p_1}\,\mathrm{GPa}$, $K_2=10^{p_2}\,\mathrm{GPa}$, and $G_2=10^{p_3}\,\mathrm{GPa}$. The anisotropy coefficient $a\in[0,1)$, the exponents $p_1,p_2,p_3\in[-3,3)$ and four angular parameters are sampled using an eight-dimensional scrambled Sobol sequence. Here, the indices $1$ and $2$ refer to the fiber and the matrix, respectively. Figure~\ref{fig:problem_setup_elastic_contrast} shows the sampled stiffness parameters in the $(p_2,p_3)$ plane, colored by the anisotropy parameter $a$. The isotropic constituents of the inelastic data, normalized by the bulk modulus of the fiber, correspond to $p_1=-0.162$, $p_2=-1.389$, $p_3=-1.725$ and $a=0$, see the marker in Fig.~\ref{fig:problem_setup_elastic_contrast}, and lie within the sampled range. The corresponding effective stiffness tensors of the fixed microstructure provide the elastic homogenization targets. The samples are split into $800$ training, $100$ validation and $100$ test cases. This elastic identification family differs from the fixed isotropic constituent properties used for the inelastic reference trajectories. As the effective stiffness is positively homogeneous of degree one in the constituent stiffnesses, only the stiffness contrasts matter, and the normalization $K_1=1\,\mathrm{GPa}$ is no restriction.
\paragraph{Inelastic data.}
For inelastic data generation, the polymer matrix obeys von Mises plasticity with linear isotropic hardening, Young's modulus of $2.1\,\mathrm{GPa}$, Poisson's ratio of $0.3$, a yield stress of $29\,\mathrm{MPa}$, and a hardening modulus of $139\,\mathrm{MPa}$. The fiber phase is linear elastic with Young's modulus $E_f=72\,\mathrm{GPa}$ and Poisson's ratio $\nu_f=0.22$. The terminal macroscopic strains were constructed from $1000$ symmetric tensors $\fE_s$ with independent, standard normally distributed components by first removing their volumetric part,
\begin{equation}
	\dev\fE_s := \fE_s-\frac{1}{3}\,\tr(\fE_s)\,\fI,
\end{equation}
and subsequently normalizing the resulting deviatoric direction as
\begin{equation}
	\bar{\feps}^{\max}_s := R^{\max} \frac{\dev\fE_s}{\|\dev\fE_s\|_F}, \qquad R^{\max}=0.05, \qquad s=1,\ldots,1000.
\end{equation}
Consequently, all terminal strains satisfy the conditions $\tr(\bar{\feps}^{\max}_s)=0$ and $\|\bar{\feps}^{\max}_s\|_F=R^{\max}$. For each terminal strain, the FFT-based solver was subjected to the monotonically increasing proportional history
\begin{equation}
	\bar{\feps}_s(\kappa) := \kappa\,\bar{\feps}^{\max}_s, \qquad \kappa\in[0,1],
\end{equation}
in terms of the load factor $\kappa$, discretized by $N_t = 10$ time steps, and provided the corresponding effective-stress trajectory. One of the $1000$ FFT computations was invalid because the norm of the final effective stress was below $10^{-6}\,\mathrm{GPa}$, and it was discarded, so that $999$ inelastic trajectories were available.

To visualize the sampling, we projected the terminal deviatoric strain onto its orthonormal deviatoric Mandel basis~\cite{krause2025tensorial} and obtained the coordinates
\begin{subequations}
	\label{eq:problem-setup-projected-coordinates}
	\begin{align}
		a_{1,s} &:=\frac{\bar\feps^{\max}_{11,s}-\bar\feps^{\max}_{22,s}}{\sqrt{2}},\\
		a_{2,s} &:=\frac{\bar\feps^{\max}_{11,s}+\bar\feps^{\max}_{22,s}-2\,\bar\feps^{\max}_{33,s}}{\sqrt{6}},\\
		a_{3,s} &:=\sqrt{2}\,\bar\varepsilon^{\max}_{12,s},\\
		a_{4,s} &:=\sqrt{2}\,\bar\varepsilon^{\max}_{13,s},\\
		a_{5,s} &:=\sqrt{2}\,\bar\varepsilon^{\max}_{23,s}.
	\end{align}
\end{subequations}
Figure~\ref{fig:problem_setup_inelastic_sampling} shows the sample distribution on the $(a_1,a_2)$ plane. In the following, the superscripts $e$ and $i$ distinguish elastic and inelastic training, respectively, while the subscript denotes the network rank. For example, \DMN{e}{5} and \DMN{i}{5} denote the corresponding rank-$5$ models. A superscript attached to \DMN{i}{} indicates the number of available inelastic trajectories.

\paragraph{Training configuration.}
The DMNs were implemented in JAX~\cite{jax2018github} and optimized using Adam~\cite{kingma2014adam}. A constant learning rate of $10^{-3}$ was employed. The eDMN was trained using mini-batches of $64$ elastic stiffness samples, whereas the iDMN used mini-batches of eight complete stress-strain trajectories and the discrete-adjoint gradient derived in Section~\ref{sec:inelastic-training-adjoint}. Each iDMN optimization was initialized by a one-sample elastic primer consisting of $1000$ Adam updates with learning rate $10^{-2}$.

During the main optimization, the validation loss was evaluated every $25$ parameter updates. Checkpoints were written every $100$ parameter updates. The best parameters were retained whenever the validation loss improved by at least $10^{-6}$, and training was terminated after $20$ consecutive validation evaluations without such an improvement, subject to a maximum of $10000$ parameter updates. Five independent runs were performed for each eDMN rank and three for each iDMN rank and training-set size. The elastic data set was split into $800$ training, $100$ validation and $100$ test samples, whereas the $999$ inelastic trajectories were split into $899$ training, $49$ validation and $51$ test trajectories. Table~\ref{tab:dmn_config} summarizes the configuration of the elastic and inelastic DMN training. It should be noted that the dataset of iDMN$^{128}$,iDMN$^{256}$, iDMN$^{512}$ are fixed and nested from that of iDMN$^{\textrm{full}}$. Moreover, the validation and test trajectories are retained identical across ranks, training-set sizes and initialization seeds.
	
	\begin{table}[ht]
		\centering
		\caption{Data-set and optimization configurations used for elastic and inelastic DMN training.}
		\label{tab:dmn_config}
		\renewcommand{\arraystretch}{1.15}
		\setlength{\tabcolsep}{5pt}
		\begin{tabular}{lccccc}
			\toprule
			& eDMN
			& iDMN$^{128}$
			& iDMN$^{256}$
			& iDMN$^{512}$
			& iDMN$^{\textrm{full}}$ \\
			\midrule
			Independent runs       & 5     & 3   & 3   & 3   & 3   \\
			Training samples       & 800   & 128 & 256 & 512 & 899 \\
			Mini-batch size        & 64    & 8   & 8   & 8   & 8   \\
			Learning rate          & $10^{-3}$ & $10^{-3}$ & $10^{-3}$ & $10^{-3}$ & $10^{-3}$ \\
			Validation interval    & 25    & 25  & 25  & 25  & 25  \\
			Checkpoint interval    & 100   & 100 & 100 & 100 & 100 \\
			Early-stop patience    & 20    & 20  & 20  & 20  & 20  \\
			Maximum updates        & 10000 & 10000 & 10000 & 10000 & 10000 \\
			\bottomrule
		\end{tabular}
	\end{table}
\paragraph{Error measures.}
For a test sample $s$ and an independently trained model $r$, the trajectory-relative effective-stress error is defined as
\begin{equation}
	\label{eq:elastic_training_e_s}
	e_s^{(r)} := \frac{\|\bar{\fsigma}^{\mathrm{pred},(r)}_s-\bar{\fsigma}^{\mathrm{FFT}}_s\|_F} {\|\bar{\fsigma}^{\mathrm{FFT}}_s\|_F}.
\end{equation}
For each training run, we first average this quantity over the $N_s$ test samples,
\begin{equation}
	\bar e^{(r)} := \frac{1}{N_s}\sum_{s=1}^{N_s}e_s^{(r)}.
\end{equation}
Because each network configuration was trained repeatedly, the statistics reported below characterize the variability of this mean test error across the $N_r$ independent runs,
\begin{equation}
	\label{eq:elastic_training_testing_errors}
	e_{\max} := \max_{1\leq r\leq N_r} \bar e^{(r)}, \qquad e_{\mathrm{mean}} := \frac{1}{N_r}\sum_{r=1}^{N_r}\bar e^{(r)}, \qquad e_{\mathrm{std}} := \sqrt{\frac{1}{N_r-1}\sum_{r=1}^{N_r}\qty(\bar e^{(r)}-e_{\mathrm{mean}})^2}.
\end{equation}

\subsection{Training results}
\label{sec:examples-results}
\subsubsection{Training}
\label{sec:examples-results-training}
\begin{figure}[!ht]
		\begin{center}
			\begin{subfigure}{\textwidth}
				\centering
				\includegraphics[height=0.022\textheight]{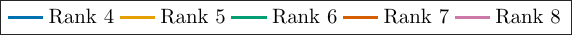}
			\end{subfigure}
		\end{center}
		\centering
		\begin{subfigure}{0.48\textwidth}
			\includegraphics[width=\linewidth]{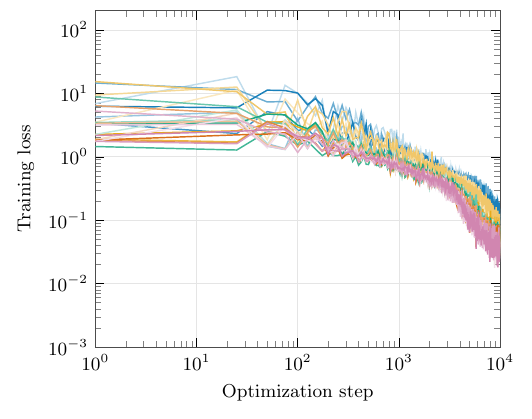}
			\caption{eDMN: training}
			\label{fig:eDMN_training_history}
		\end{subfigure}
		\hfill
		\begin{subfigure}{0.48\textwidth}
			\includegraphics[width=\linewidth]{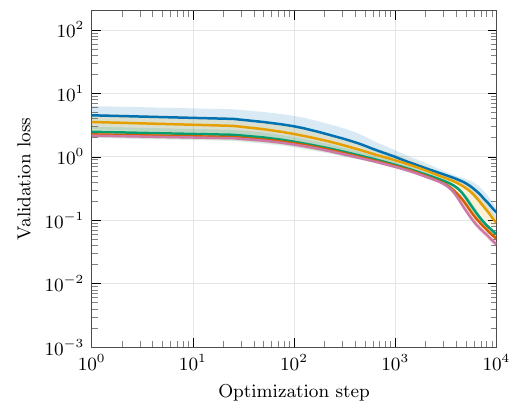}
			\caption{eDMN: validation}
			\label{fig:eDMN_validation_history}
		\end{subfigure}			
		\begin{subfigure}{0.48\textwidth}
			\includegraphics[width=\linewidth]{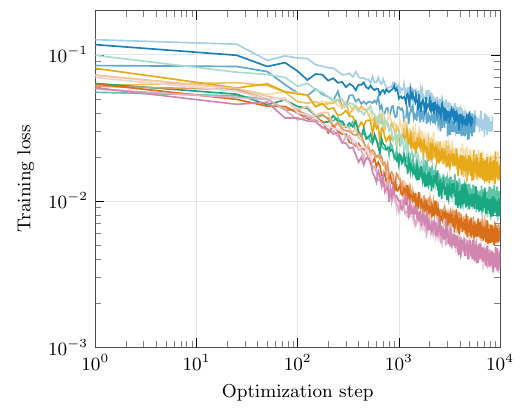}
			\caption{iDMN$^{\textrm{full}}$: training}
			\label{fig:iDMN_training_history}
		\end{subfigure}
		\hfill
		\begin{subfigure}{0.48\textwidth}
			\includegraphics[width=\linewidth]{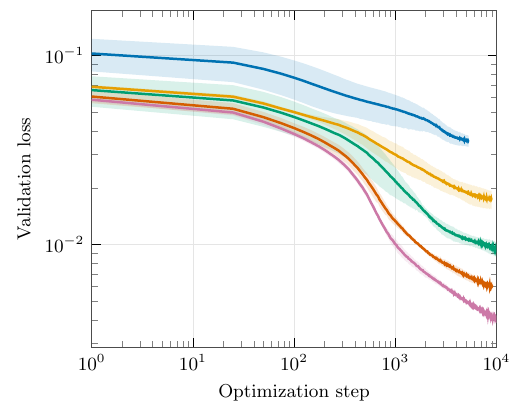}
			\caption{iDMN$^{\textrm{full}}$: validation}
			\label{fig:iDMN_validation_history}
		\end{subfigure}			
		\caption{Training and validation loss histories of the \DMN{e}{} and the iDMN$^{\textrm{full}}$. Colors denote the network rank. For the training histories, shades of each color distinguish the independent runs with different initializations. For the validation histories, solid lines denote the mean and the shaded regions the standard deviation over repeated runs.}
		\label{fig:training_history}
	\end{figure}

Figure~\ref{fig:training_history} shows the training and validation histories for the \DMN{e}{} and the iDMN$^{\textrm{full}}$. The other inelastic training-set sizes show a similar behavior. Because elastic and inelastic training minimize different objectives in Eqs.~\eqref{eq:elastic_training_loss_function_DMN} and \eqref{eq:inelastic_training_loss_function}, respectively, their absolute loss values are not directly comparable and should be interpreted only within each training strategy. In both cases, increasing the network rank generally lowers the converged loss, indicating that the additional material points and internal nodes improve the representation of the microscopic homogenizer. The higher loss and earlier plateau observed for the lowest rank considered here are consistent with insufficient network capacity.

The inelastic optimization was initialized by the elastic warm start described above, which provides a mechanically informed initial parameter set at low cost. Consequently, the initial \DMN{i}{} losses are generally below unity, as seen in Fig.~\ref{fig:iDMN_training_history}. The relative merits of elastic and inelastic training are therefore assessed on the common inelastic test set below rather than through their different training objectives.		
\subsubsection{Testing}
\label{sec:examples-results-testing}
\begin{figure}[!ht]
	\centering
	\includegraphics[width=0.88\textwidth]{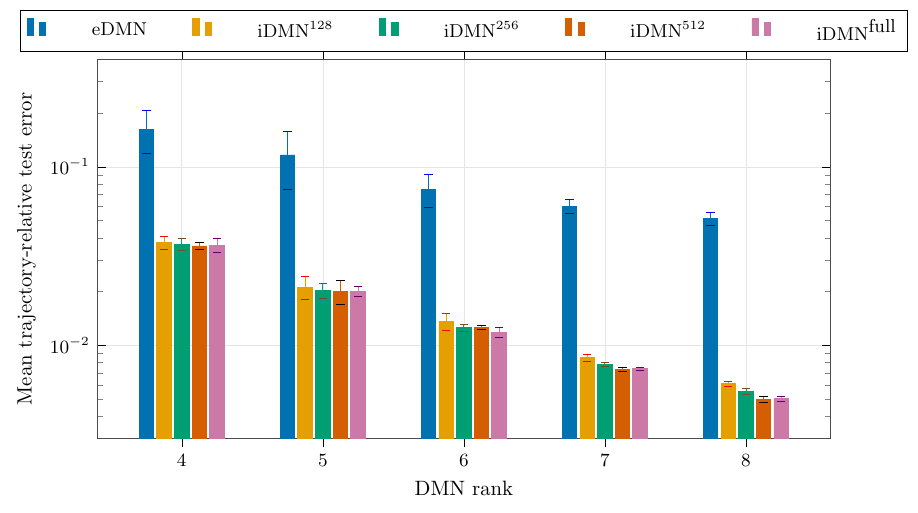}
	\caption{Mean trajectory-relative effective-stress error on the inelastic test set as a function of network rank for the eDMN and the four iDMN training-set sizes.}
	\label{fig:idmn_rank_sensitivity}
\end{figure}

The trained models were evaluated on a common inelastic test set of unseen monotonically increasing proportional strain paths. For each independently trained model, the trajectory-relative error in Eq.~\eqref{eq:elastic_training_e_s} was averaged over the test samples. Figure~\ref{fig:idmn_rank_sensitivity} and Table~\ref{tab:errors_combined_sigeff_testdata} report the resulting statistics over the repeated training runs.
\input{Gen2_results/ortho_revision/nonlinear_test/tex/iDMN_study/table_iDMN_B_rank_sensitivity_upgraded.tex}

For the eDMN, the mean test error decreases from $16.28\%$ at rank $4$ to $5.12\%$ at rank $8$. The scatter over the independent training runs is large at low rank, with standard deviations of $4.33\%$ and $4.23\%$ at the ranks $4$ and $5$. At every investigated rank, inelastic training yields a substantially lower error. Already at rank $4$, the iDMNs achieve mean errors of $3.60\%$--$3.75\%$, below the $5.12\%$ of the \DMN{e}{8}, while the \DMN{i}{8} reach $0.50\%$--$0.61\%$. Hence, already the iDMN$_{4}^{128}$ outperforms the \DMN{e}{8} despite using only $128$ inelastic training trajectories.

Increasing the number of inelastic trajectories yields further improvement, but with diminishing returns. At rank $8$, the mean error changes only from $0.61\%$ for iDMN$^{128}$ to $0.50\%$ for iDMN$^{512}$ and iDMN$^{\textrm{full}}$. The corresponding standard deviations are all below $0.02\%$. The iDMN$^{256}$, iDMN$^{512}$, and iDMN$^{\text{full}}$ models are already nearly indistinguishable at ranks $7$--$8$. These results separate the effects of model capacity and training-set size. Increasing the network rank consistently improves the accuracy, whereas beyond a moderate number of inelastic trajectories the sensitivity to the data-set size is weak for the present loading distribution. Thus, for the proportional-loading test considered here, network capacity appears to become the dominant limitation under the present training protocol once a moderate inelastic training set is available.
\subsubsection{Computational cost}
\label{sec:examples-results-cost}
This subsection examines the three components of computational cost: Generation of the elastic and inelastic reference data, training of the eDMN and iDMN models, and online acceleration relative to the full-field FFT-based solver.

\paragraph{Data generation}
Table~\ref{tab:dataset_generation_time} reports the offline cost of the full-field simulations used to construct the training targets. All full-field computations were carried out on a workstation with AMD EPYC 9354 processors. Each elastic sample required six linear FFT-based homogenization problems, each of which took about $78.4\,\mathrm{s}$ on $12$ threads, i.e., approximately $1.57$ CPU-hours per sample and $1568.34$ CPU-hours for the $1000$ samples. Each inelastic trajectory took $173.7\,\mathrm{s} \pm 16.0\,\mathrm{s}$ wall-clock time on six threads, i.e., approximately $0.29$ CPU-hours, and the $999$ inelastic trajectories required $289.27$ CPU-hours in total, from which we deduced the generation times of $128$, $256$ and $512$ trajectories. Thus, the generation of the inelastic data consumed less than a fifth of the CPU time of the elastic data set of comparable size. This comparison should be interpreted in view of the different targets: An elastic sample contains a complete effective stiffness tensor and thus requires several independent linear homogenization problems, whereas an inelastic sample contains the response to one prescribed proportional loading trajectory. Moreover, the FFT-based solver benefits from extrapolation in the inelastic case: The converged solutions of the preceding load steps provide an accurate initial guess for the current load step~\cite{schneider2021review}, so that few iterations per load step suffice. In contrast, the six load cases of an elastic sample are independent of each other, and no such extrapolation is available. Hence, the lower cost does not imply that inelastic constitutive integration is intrinsically less expensive than linear homogenization.
\begin{table}[!ht]
		\centering
		\caption{CPU time required to generate the elastic and inelastic training data sets.}
		\label{tab:dataset_generation_time}
		\begin{tabular}{crr}
			\toprule
			Dataset & Samples & Time (CPU-hours) \\
			\midrule
			Elastic training & 1000 & 1568.34 \\
			\midrule
			Inelastic training & 128 & 37.06 \\
			& 256 & 74.13 \\
			& 512 & 148.25 \\
			& 999 & 289.27 \\
			\bottomrule
		\end{tabular}
	\end{table}
For the inelastic training, the reference data were generated once and reused for every network rank, so this part of the offline cost depends on the selected data-set size but not on the DMN architecture.

\begin{figure}[!ht]
	\centering
	\includegraphics[width=0.82\textwidth]{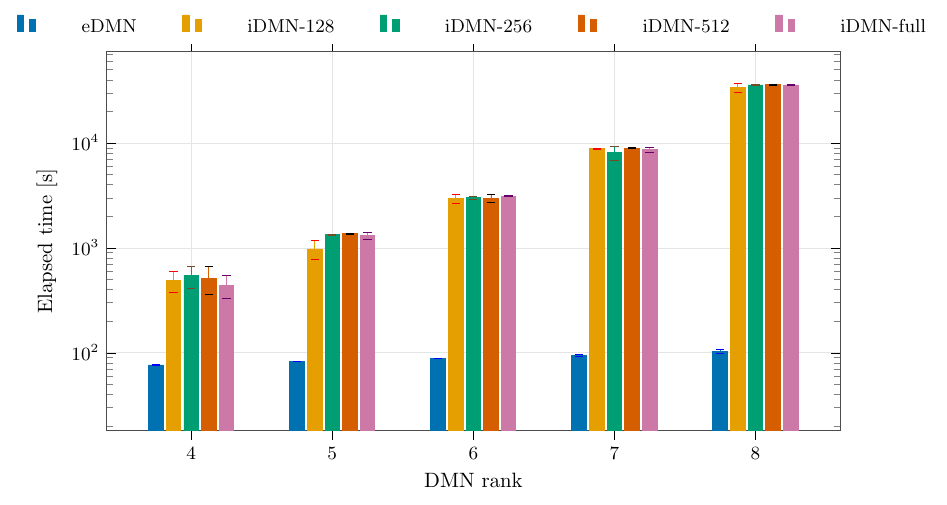}
	\caption{Training time for various network ranks and data-set sizes of the eDMN and iDMN models. Error bars represent the variability over repeated runs.}
	\label{fig:total_time_seconds_comparison}
\end{figure}

\paragraph{Training eDMN and iDMN}
The training times of the eDMN and iDMN are reported in Tab.~\ref{tab:elapsed_seconds_combined} and plotted in Fig.~\ref{fig:total_time_seconds_comparison}. Elastic training remains inexpensive and only weakly dependent on rank: The mean elapsed time increases from $76.47\,\mathrm{s}$ at rank $4$ to $102.83\,\mathrm{s}$ at rank $8$. The cost of acquiring the eDMN stems almost exclusively from the data generation. On the other hand, inelastic training is substantially more expensive because each update integrates and differentiates the complete history-dependent constitutive problem. Its mean elapsed time rises from approximately $439$--$540\,\mathrm{s}$ at rank $4$ to $3.36\times10^4$--$3.59\times10^4\,\mathrm{s}$ at rank $8$, corresponding to roughly $9$--$10$ hours for the deepest networks. The training was carried out on a single NVIDIA H100 PCIe GPU of the supercomputer amplitUDE, whereas the reference data were generated on CPUs. As these resources are not directly comparable, we report the cost of data generation in CPU-hours and the cost of training in GPU wall-clock time separately. For the iDMN$^{128}$, for instance, the data generation required $37.06$ CPU-hours, about one fortieth of the $1568.34$ CPU-hours for the elastic data set, whereas the training at rank $6$ took about $0.8$ GPU-hours compared with about one and a half minutes for the eDMN.
\input{Gen2_results/ortho_revision/computational_cost/tex/iDMN_study/table_elapsed_seconds_comparison.tex}

At a fixed rank, the training time of the iDMN hardly depends on the size of the inelastic data set. This observation is a consequence of the training protocol, which prescribes the mini-batch size, the validation interval and the maximum number of parameter updates, and does not imply that processing an epoch is independent of the data-set size. The cost increase is therefore primarily associated with the size of the nonlinear network problem and its history-dependent adjoint solve, rather than scaling proportionally with the number of available trajectories. This result complements the accuracy study: Increasing the inelastic data set beyond a moderate size produces diminishing accuracy gains, while increasing rank rapidly raises the offline cost. The favorable region is consequently at intermediate rank, where inelastic training captures most of the accuracy improvement without the extreme rank-$8$ training expense.

\paragraph{Comparison with FFT-based solver.}
For online evaluation, the trained parameters and consequently the kinematic operator $D$ were held fixed. The Fortran runtime is therefore independent of whether these material network parameters were obtained from elastic or inelastic training. At each time step, the material-point constitutive states were updated and the nonlinear global equilibrium equation~\eqref{eq:setup-flattened-residual} was solved, using PETSc with a Newton line-search method~\cite{petsc-web-page}. The global residual tolerance is chosen to be $10^{-10}$. The consistent reduced tangent was applied matrix-free, and the resulting linearized system was solved by the conjugate-gradient method. A fixed elastic-reference tangent
\begin{equation}
	K_{\mathrm{el}}:=\sum_i w_i D_i^T \C_i^{\mathrm{el}}D_i,
\end{equation}
in terms of the elastic stiffnesses $\C_i^{\mathrm{el}}$ and the blocks $D_i$ of the operator $D$ associated with the material points was assembled and Cholesky-factorized once and subsequently reused as a preconditioner throughout the nonlinear trajectory. The online prediction is performed on one CPU of a desktop computer (Intel Core Ultra 7 265) with Intel ifx 2025.3.2 compiler, using PETSc version 3.25.4. For each model rank, we perform 5 simulation and compute the mean time.

The online efficiency was assessed from the wall-clock time required to evaluate one prescribed macroscopic strain trajectory from the test set. The FFT-based reference solver required $182.968\,\mathrm{s}\pm11.159\,\mathrm{s}$. Table~\ref{tab:speedup} reports the resulting speed-ups for the \DMN{e}{} and iDMN$^{128}$ models evaluated with the same Fortran solver.
\begin{table}[h!]
	\centering
	\caption{Speed-up of the Fortran DMN solver relative to the FFT-based full-order solver, computed from the corresponding mean wall-clock runtimes.}
	\label{tab:speedup}
		\begin{tabular}{lccccc}
			\toprule
			Model & Rank 4 & Rank 5 & Rank 6 & Rank 7 & Rank 8 \\
			\midrule
			\DMN{e}{} & $1.43\times10^{5}$ & $5.14\times10^{4}$ & $1.14\times10^{4}$ & $1.74\times10^{3}$ & $2.57\times10^{2}$ \\
			iDMN$^{128}$ & $1.45\times10^{5}$ & $3.78\times10^{4}$ & $1.32\times10^{4}$ & $2.25\times10^{3}$ & $3.36\times10^{2}$ \\
			\bottomrule
		\end{tabular}
\end{table} 
The online acceleration decreases strongly with increasing network rank because both the number of material points and the number of equilibrium unknowns grow with network depth. The speed-up decreases from approximately $1.5\times10^{5}$--$2.2\times10^{5}$ at rank $4$ to approximately $3.3\times10^{2}$--$3.4\times10^{2}$ at rank $8$, while ranks $5$--$7$ provide intermediate accelerations of order $10^{3}$--$10^{4}$. Thus, even for the largest network considered here, the reduced DMN remains more than two orders of magnitude faster than the FFT-based full-field solver.
At a fixed rank, the \DMN{e}{} and iDMN$^{128}$ show a comparable online cost. This is expected because both were evaluated through the same frozen material-network formulation and differ only in the identified network parameters. The training strategy therefore primarily affects the offline identification cost and predictive accuracy rather than the complexity of the online solver.

\subsection{Online prediction} 
\label{sec:examples-Online}
\begin{figure}[!h]
	\begin{center}
		\begin{subfigure}{\textwidth}
			\centering
			\includegraphics[height=0.022\textheight]{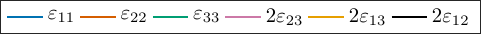}
		\end{subfigure}
	\end{center}
	\centering
	\begin{subfigure}[t]{0.32\textwidth}
		\includegraphics[width=\linewidth]{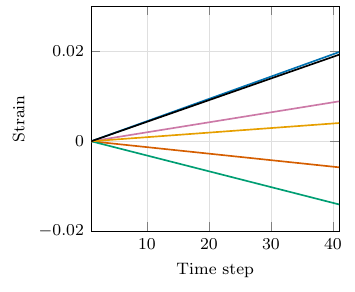}
		\caption{Proportional ramp-up}
	\end{subfigure}
	\hfill
	\begin{subfigure}[t]{0.32\textwidth}
		\includegraphics[width=\linewidth]{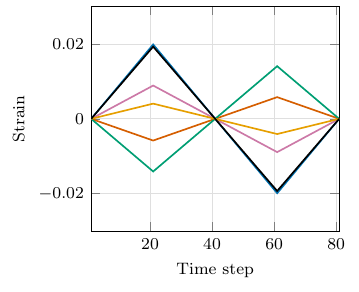}
		\caption{Cyclic proportional}
	\end{subfigure}	
	\hfill		
	\begin{subfigure}[t]{0.32\textwidth}
		\includegraphics[width=\linewidth]{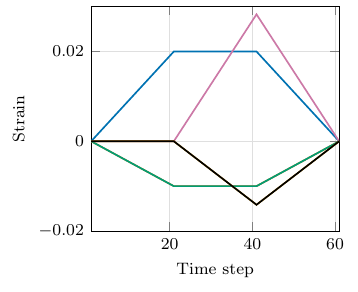}
		\caption{Non-proportional}
	\end{subfigure}
	 \caption{Strain paths for online extrapolation.}
	 \label{fig:Applied_strains}		
\end{figure}

\begin{figure}[!ht]
	\centering
	\includegraphics[width=0.88\textwidth]{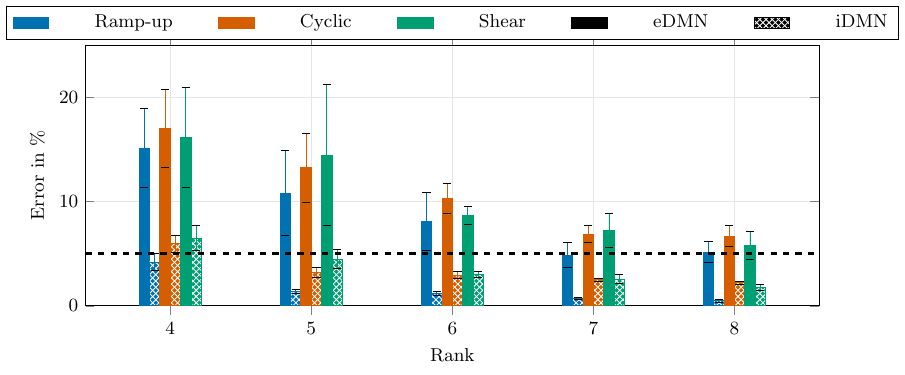}
	\caption{Trajectory-relative effective-stress errors of the \DMN{e}{} and iDMN$_{}^{\textrm{full}}$ for the three unseen loading paths. The dashed line denotes the $5\%$ reference error threshold.}
	\label{fig:dmn_ood_trajectory_error}
\end{figure}

This section assesses the predictive capability of the trained \DMN{e}{} and \DMN{i}{} models for loading histories that differ from the training data. Specifically, the material networks were evaluated on three unseen loading histories: a monotonic proportional ramp-up path with a finer temporal discretization, a proportional cyclic path, and a non-proportional path. The ramp-up path belongs to the same family of paths as the training data and serves as a transfer test with respect to the temporal resolution, whereas the cyclic and the non-proportional paths are out of distribution (OOD). These tests represent progressively more demanding departures from the inelastic training distribution. The ramp-up case preserves the monotonic and proportional character of the training paths, while introducing an unseen loading direction and a finer temporal discretization. The cyclic case additionally introduces unloading, reverse loading, and reloading. Finally, the non-proportional case changes both the loading direction and the loading history throughout the trajectory. All three paths were discretized using smaller time increments than those employed in the training dataset of \DMN{i}{}, with $N_t = 40$, $80$ and $60$ time steps, respectively, as illustrated in Fig.~\ref{fig:Applied_strains}.
The effective-stress trajectories predicted by the trained material networks were compared with reference solutions obtained using the FFT-based micromechanical solver. We show in Appendix~\ref{app:online_prediction_effective_stress} the predictions from \DMN{e}{8} and \DMN{i}{8} together with the component-wise relative error. The primary error measure is the trajectory-relative norm defined in Eq.~\eqref{eq:elastic_training_e_s}, applied to each of the three loading paths. Figure~\ref{fig:dmn_ood_trajectory_error} and Table~\ref{tab:dmn_ood_trajectory_error} report this global trajectory error as a function of network rank.
\input{Gen2_results/ortho_revision/ood/analysis/table_ood_gajek_all_seeds.tex}

Three observations follow from Fig.~\ref{fig:dmn_ood_trajectory_error}. First, the mean errors of the \DMN{i}{8} remain below the $5\%$ trajectory-error threshold for all three unseen loading paths, whereas those of the \DMN{e}{8} exceed it on all three paths. The inelastically trained DMN reduces the mean ramp-up error from $5.16\%$ to $0.48\%$, the cyclic error from $6.68\%$ to $2.21\%$, and the non-proportional error from $5.79\%$ to $1.77\%$. The iDMN$^{\textrm{full}}$ stays below the $5\%$ threshold on all three paths from rank $5$ onward, and the \DMN{i}{5} is more accurate than the \DMN{e}{8} on every path. Moreover, the scatter over the independent training runs is considerably smaller for the iDMN than for the eDMN. Since the two models have the same network rank, this improvement results from the training strategy rather than from increased network capacity. Second, the accuracy gained from inelastic training is retained beyond the monotonically increasing proportional trajectories used for identification. The smallest error is obtained for the ramp-up path, which preserves the proportional character of the training trajectories but changes the loading direction and temporal discretization. The advantage remains substantial for cyclic loading, which introduces unloading and reverse loading, and for non-proportional loading, in which the strain direction evolves throughout the trajectory. Thus, the inelastic identification improves not only interpolation within the training setting but also transfer to qualitatively different loading histories. Third, the larger errors for the cyclic and non-proportional paths relative to the ramp-up case indicate that the trained DMN remains sensitive to the distribution of loading histories used for identification. The present inelastic training data contain monotonically increasing proportional trajectories only. Consequently, increasing the accuracy for more general loading histories may require greater diversity of the training trajectories rather than merely additional samples of the same proportional-loading family.

\subsection{Transfer to perturbed constituent properties}
\label{sec:material-transfer}
The loading-path tests above vary the macroscopic history while retaining the constituent properties used for inelastic identification. We next examine the performance of the iDMN for different material properties. The cyclic loading path was retained because it already probes unloading, reverse loading, and reloading, while the trained network parameters were kept fixed. Starting from the nominal matrix parameters, i.e., the Young's modulus $E_m=2.1\,\mathrm{GPa}$, the Poisson's ratio $\nu_m=0.30$, the yield stress $\sigma_y=29\,\mathrm{MPa}$, and the hardening modulus $H=139\,\mathrm{MPa}$, we perturbed one parameter at a time by $\pm10\%$. The fiber remains linear elastic with the Young's modulus $E_f=72\,\mathrm{GPa}$ and the Poisson's ratio $\nu_f=0.22$. Together with the nominal case, this procedure results in nine material scenarios. We again relied on the FFT-based solver to obtain the reference trajectories, which were then used to determine the trajectory-relative norm in Eq.~\eqref{eq:elastic_training_e_s}.

This experiment also compares the elastic and inelastic identification budgets. The eDMN was identified from $800$ elastic stiffness samples spanning a range of constituent properties, whereas the iDMN$^{128}$ used only $128$ inelastic trajectories at the nominal material parameters. The stiffness contrasts of the nominal matrix and of all eight perturbations lie within the sampled range of the elastic data, see Section~\ref{sec:examples-setting}. Hence, both the eDMN and the iDMN$^{128}$ are applied in the vicinity of their training data. %	We do not reproduce the material-informed elastic sampling of Gajek et al.~\cite{gajek2023material}: That construction exploits the constitutive structure of linear viscoelasticity to concentrate elastic samples in the stiffness region relevant to the subsequent response. Instead, we test material transfer directly by changing the constituent parameters in the full-field and reduced models.
Table~\ref{tab:material_transfer_edmn_idmn128} compares the eDMN and iDMN$^{128}$ over all ranks and material scenarios. The complete $25$-model sweep, including iDMN$^{256}$, iDMN$^{512}$, and iDMN$^{\mathrm{full}}$, is reported in Appendix~\ref{app:material_transfer}.

\input{Gen2_results/ortho_revision/material_transfer/R5_run/table7_edmn_idmn128.tex}

\begin{figure}[h!]
	\begin{center}
		\begin{subfigure}{\textwidth}
			\centering
			\includegraphics[height=0.022\textheight]{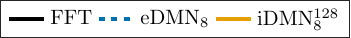}
		\end{subfigure}
	\end{center}
	\centering
	\begin{subfigure}{0.24\textwidth}\includegraphics[width=\linewidth]{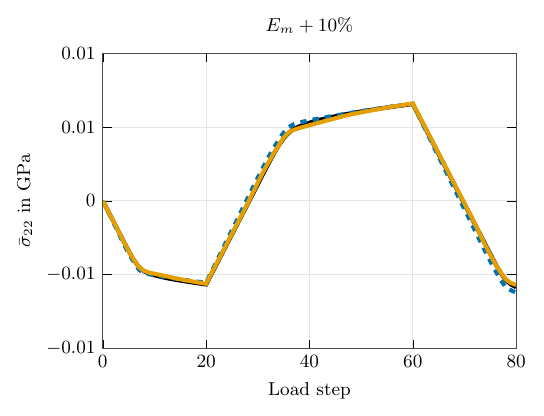}\caption{$E^{+}$}\end{subfigure}
	\begin{subfigure}{0.24\textwidth}\includegraphics[width=\linewidth]{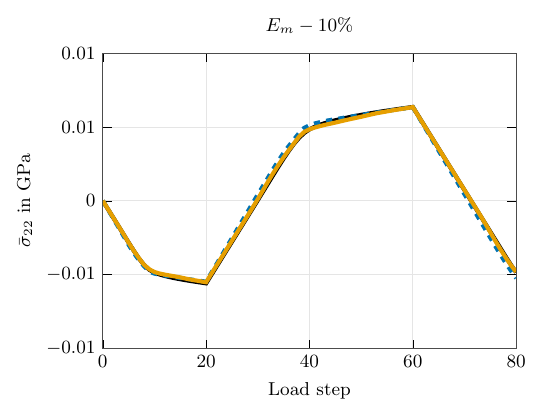}\caption{$E^{-}$}\end{subfigure}
	\begin{subfigure}{0.24\textwidth}\includegraphics[width=\linewidth]{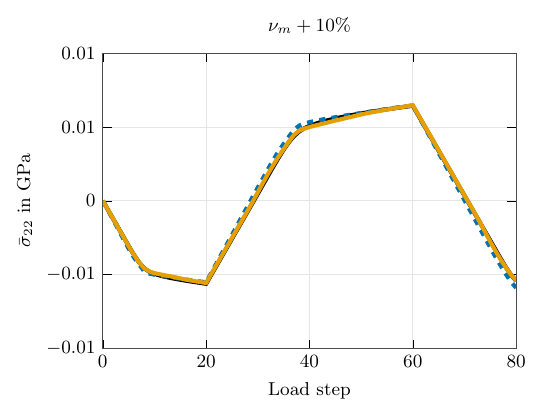}\caption{$\nu^{+}$}\end{subfigure} 		
	\begin{subfigure}{0.24\textwidth}\includegraphics[width=\linewidth]{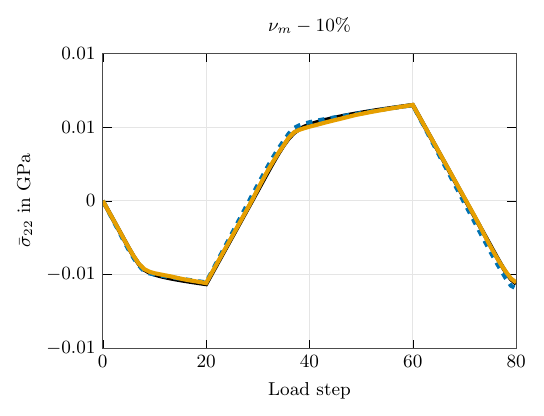}\caption{$\nu^{-}$}\end{subfigure}
	
	\begin{subfigure}{0.24\textwidth}\includegraphics[width=\linewidth]{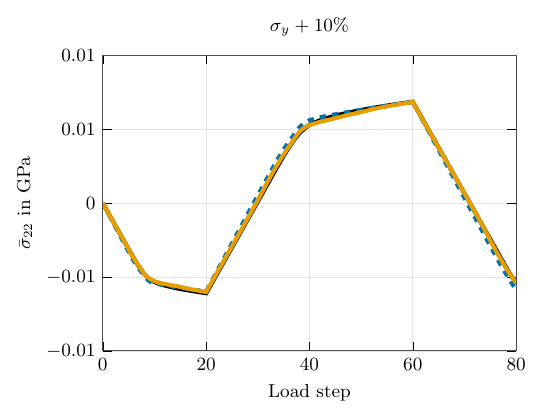}\caption{$\sigma_y^{+}$}\end{subfigure}
	\begin{subfigure}{0.24\textwidth}\includegraphics[width=\linewidth]{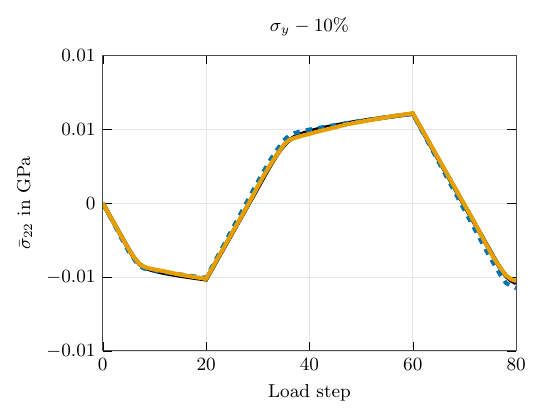}\caption{$\sigma_y^{-}$}\end{subfigure}
	\begin{subfigure}{0.24\textwidth}\includegraphics[width=\linewidth]{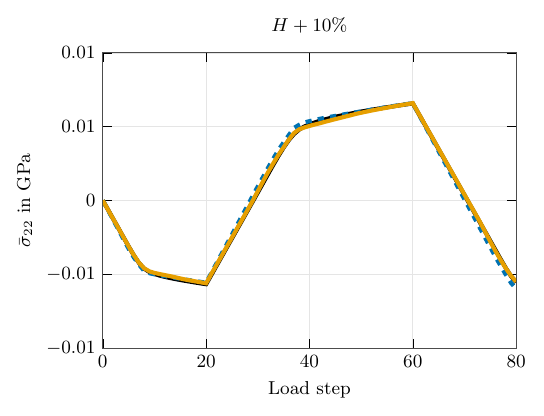}\caption{$H^{+}$}\end{subfigure} 		
	\begin{subfigure}{0.24\textwidth}\includegraphics[width=\linewidth]{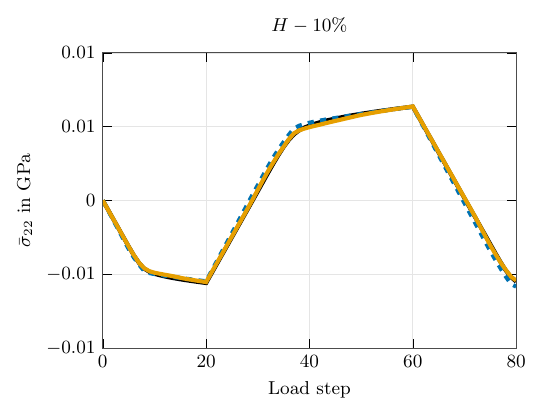}\caption{$H^{-}$}\end{subfigure}
	%note to myself: seed-0 runs, except eDMN7 seed-1.
	\caption{Predicted effective stress $\bar{\sigma}_{22}$ for perturbed material parameters of the matrix, shown for one training run per model.}
	\label{fig:material_transfer_sigma22}
\end{figure}

For iDMN$^{128}$, the inelastically identified model has lower error than the eDMN in every one of the $45$ same-rank comparisons in Table~\ref{tab:material_transfer_edmn_idmn128}. At rank $8$, its nominal, mean perturbed, and worst perturbed errors are $2.27\%$, $2.28\%$, and $2.44\%$, respectively, compared with $6.68\%$, $6.70\%$, and $6.83\%$ for the eDMN, where all values refer to the mean over the independent training runs. For this cyclic transfer test, even the iDMN$_{4}^{128}$ remains more accurate than the \DMN{e}{8} for all nine material cases. These observations also hold true for other inelastic training-set sizes.  
	
For a closer look, Figure~\ref{fig:material_transfer_sigma22} presents the trajectory of the effective stress component $\bar\sigma_{22}$ for the eight perturbed material properties. The iDMN$^{128}$ trajectory remains close to the FFT reference as the material parameters change, whereas the eDMN shows larger deviations, as indicated by the errors in Table~\ref{tab:material_transfer_edmn_idmn128}.

The experiment therefore shows that the accuracy gained by inelastic identification is not confined to the constituent parameters used during training. Within the investigated $\pm10\%$ neighborhood, it is retained under changes of elastic modulus, Poisson ratio, yield stress, and isotropic hardening modulus. This is a local material-transfer result around the present $J_2$ constitutive law and does not establish transfer to arbitrary constituent models or parameter ranges.

\section{Conclusions}
\label{sec:conclusions}
In this work, we identified deep material networks directly from inelastic stress-strain trajectories. To this end, we formulated the training as a history-dependent constrained optimization problem and derived a path-consistent discrete adjoint, which provides the complete sensitivities of the trajectory loss with one backward solve of the linearized state equations per time step, independently of the number of trainable parameters. For a short-fiber composite with an elastoplastic matrix, the inelastically trained networks turned out to be more accurate than their elastically trained counterparts, showed a smaller scatter over independent training runs and required less CPU time for the generation of their training data. In the following, we discuss what can be learned from these findings.

Elastic training is the standard identification strategy for DMNs, and for good reasons. It relies on efficient linear solvers, it is robust, and every trained network inherits thermodynamic consistency from its constituents. Its theoretical foundation for inelastic applications, however, is narrower than its practical success might suggest. Gajek et al.~\cite{gajek2020micromechanics} showed that the effective response of a material with vanishing memory is determined to first order by the linear elastic homogenization function. Thus, the transfer from elastic training to inelastic behavior is established only asymptotically and only for this class of materials. Moreover, the expressivity of DMNs is open even in linear elasticity, as it is unknown whether the set $G$ of effective stiffnesses realizable by arbitrary microstructures of the constituents coincides with the set $G_L$ of effective stiffnesses realizable by hierarchical laminates~\cite{Milton2022,milton2021open}. Although inelastic training leaves these questions open, it permits assessing the approximation quality of DMNs for inelastic constituents directly.

In practice, it is often unclear how and with which data a DMN should be trained elastically. Dey et al.~\cite{dey2022training} observed that networks with similar elastic errors may show markedly different creep errors, and Gajek et al.~\cite{gajek2023material} improved the elastic-to-viscoelastic transfer by a sampling tailored to the Maxwell structure of the constituents. Our results point in the same direction and indicate that, for the present problem and elastic sampling, the elastic objective does not constrain the inelastic response sufficiently. At rank $8$, the test error of the eDMN amounts to $5.12\,\%$ and is an order of magnitude larger than the error of about $0.5\,\%$ of the iDMN. Similarly, on the unseen loading paths, the scatter of the eDMN error over independent training runs is up to an order of magnitude larger than that of the iDMN. Additional network capacity therefore pays off only if the training data inform the network about the history-dependent strain partitioning. Inelastic training provides this information directly, without requiring a sampling strategy tailored to the specific material class.

The higher accuracy per network level is of immediate practical relevance, because the online cost of a DMN is governed by the number $M=2^K$ of material-point evaluations, which doubles with every additional level. Keeping the network shallow is therefore essential for efficient two-scale simulations. In our study, the rank-$5$ iDMNs were more accurate than the \DMN{e}{8} on the test set and on all investigated unseen loading paths. At the same time, the online speed-up of the iDMN$_5^{128}$ over the full-field solver amounts to $3.78\times10^{4}$, which exceeds the speed-up of the \DMN{e}{8} by a factor of about $150$.

The offline cost of elastic training is easily underestimated, because the data generation dominates the identification. Every elastic sample requires six independent linear homogenization problems, whereas an inelastic trajectory is computed in a single run, which benefits from extrapolation between load steps. For the composite considered here, the inelastic data set required less than a fifth of the CPU time of the elastic one, and already $128$ trajectories sufficed for an accuracy that the eDMN did not reach at any rank. In particular, the data generation for the iDMN$_{5}^{128}$ required about one fortieth of the CPU-hours of the elastic data set, whereas its training took about a quarter of a GPU-hour. In this respect, inelastic training brings DMNs closer to classical model order reduction, where inelastic data are required anyway. For projection-based reduced-order models, the proper orthogonal decomposition of inelastic snapshots constitutes the training~\cite{yvonnet2007reduced}, and hyper-reduction techniques such as the empirical cubature method~\cite{hernandez2017dimensional,lange2024monolithic} and the empirically corrected cluster cubature~\cite{wulfinghoff2025homogenization,wulfinghoff2025e3c,wulfinghoff2026computational} select their integration points or clusters on the basis of inelastic data as well. In contrast to these methods, a DMN does not approximate the solution manifold by a linear subspace. Consequently, the DMN is not subject to the approximation barrier associated with projecting the solution manifold onto a fixed linear subspace, which is quantified by the Kolmogorov $n$-width.

The inelastically identified networks also generalize beyond their training setting. Although the training data comprised monotonically increasing proportional paths for a single set of material parameters only, the iDMN remained more accurate than the eDMN for cyclic and non-proportional loading as well as under $\pm 10\,\%$ perturbations of the elastic and plastic matrix parameters. The comparatively larger errors for cyclic and non-proportional loading suggest that future training sets should cover a greater variety of loading paths rather than more samples of the same path family.

These benefits come at the price of a considerably more involved implementation. The training itself requires only macroscopic stress trajectories, and the internal variables remain unobserved. To obtain path-consistent sensitivities, however, the dependence of the inherited constitutive state needs to be propagated from one increment to the next. For this purpose, the material routine has to provide the derivatives of the constitutive update with respect to the state of the previous time step in addition to the consistent tangent, which goes beyond the interface of standard user material routines. Furthermore, the coupled global-local adjoint system has to be solved backward in time on the stored forward trajectory, and the resulting gradients require careful verification, as the constitutive response switches between elastic and plastic branches along the loading path. These difficulties may be one reason why Nguyen and Noels~\cite{nguyen2022interaction} implemented a step-local simplification for monotonic loading paths.

In practice, we recommend using elastic information as an efficient initialization and devoting the high-fidelity data to the response regime in which the surrogate is deployed. Natural extensions of the present work concern training data with cyclic and non-proportional loading paths, other classes of inelastic constituents such as viscoplastic or damaging materials, and the deployment of inelastically trained DMNs in concurrent two-scale simulations.

%\t\newpage
\section*{Acknowledgements}
%The authors thank P.~Bhat Keelanje Srinivas for providing the elastic material-property samples. % obsolete: elastic data now from the sampling of Gajek et al.
The authors gratefully acknowledge the computing time granted by the Center for Computational Sciences and Simulation (CCSS) of the University of Duisburg-Essen and provided on the supercomputer amplitUDE (DFG project 459398823; grant ID INST 20876/423-1 FUGG) at the Center for Information and Media Services (ZIM).

\section*{Funding}
This work was funded by the European Union through the European Research Council under the Horizon Europe programme (ERC Starting Grant BeyondRVE, grant agreement No.~101040238). Views and opinions expressed are, however, those of the authors only and do not necessarily reflect those of the European Union or the European Research Council Executive Agency. Neither the European Union nor the granting authority can be held responsible for them.

\section*{CRediT authorship contribution statement}
\textbf{Binh Huy Nguyen:} Methodology, Software, Validation, Formal analysis, Investigation, Data curation, Visualization, Writing -- original draft, Writing -- review \& editing.
\textbf{Matti Schneider:} Conceptualization, Methodology, Formal analysis, Writing -- original draft, Writing -- review \& editing, Supervision, Funding acquisition.

\section*{Declaration of competing interest}
The authors declare that they have no known competing financial interests or personal relationships that could have appeared to influence the work reported in this paper.

\section*{Data availability}
Data will be made available on request.

\section*{Declaration of generative AI and AI-assisted technologies in the writing process}
During the preparation of this work, the authors used ChatGPT by OpenAI and Claude by Anthropic for linguistic revision, literature searches, and bibliographic checks, Claude Code by Anthropic for consistency checks of notation and the preparation of figures, and Codex by OpenAI to assist with code formatting. After using these tools, the authors reviewed and edited the content as needed, independently verified all mathematical arguments, references, and numerical results, and take full responsibility for the content of the publication.

\appendix
\section{Analytical Jacobian blocks}
\label{app:jacobian_blocks}
For the sake of clarity, we present the closed-form expressions of the Jacobian blocks~\eqref{eq:inelastic_training_Jacobian_blocks} in this section. We refer to Michaleris et al.~\cite{michaleris1994tangent} for more details. For the two-phase composite at hand, we compute the blocks of the elastic fiber and the elastoplastic matrix separately. Specifically, for each material point that belongs to the elastic fiber phase, the state vector $\fv_n^{\textrm{fiber}}$ stores only the stress
\begin{equation}
	\label{eq:appendix-concatenated-vn-fiber}
	\fv_n^{\textrm{fiber}} = \fsigma_n \in \R^{6},
\end{equation}
whereas for a material point governed by $J_2$ plasticity, the vector $\fv_n^{\textrm{matrix}}$ consists of the plastic strain $\feps^p_n$, the accumulated equivalent plastic strain $p_n$, the Cauchy stress $\fsigma_n$ and the accumulated plastic multiplier $\lambda_n$ at the current step $n$,
\begin{equation}
	\label{eq:appendix-concatenated-vn-matrix}
	\fv_n^{\textrm{matrix}} = \mqty[\feps^p_n & p_n & \fsigma_n & \lambda_n]^T \in \R^{14}.
\end{equation}
At each increment, the state vectors are updated separately and subsequently concatenated into the global vector
\begin{equation}
	\label{eq:appendix-concatenated-vn}
	\fv_n \in \R^{M(14+6)/2}
\end{equation}
for a DMN with $M$ material points. This distinction also enters the local residual and its partial derivatives. We denote by $\vb{I}_{r\times s}$ and $\vb{0}_{r\times s}$ the identity and zero matrices of dimensions $r \times s$, respectively. Each material point of the elastic fiber phase admits the local residual
\begin{equation}
	\label{eq:appendix-Hlocal-fiber}
	\fH_n^{\textrm{fiber}} = \fsigma_n - \fsigma_{n-1} - \C_{\textrm{fiber}}\qty(\feps_{n}-\feps_{n-1} ),
\end{equation}
such that the corresponding Jacobian blocks read
\begin{subequations}
	\label{eq:appendix-Jacobian-fiber}
	\begin{align}
		\fH^{\textrm{fiber},\fv}_n&=\vb{I}_{6\times 6}, \\
		\fH^{\textrm{fiber},\feps}_n&=-\C_{\textrm{fiber}},\\
		\fH^{\textrm{fiber},\fv,-}_n&=-\vb{I}_{6\times 6}, \\
		\fH^{\textrm{fiber},\feps,-}_n&=\C_{\textrm{fiber}}.
	\end{align}
\end{subequations}
A material point of the elastoplastic matrix may deform elastically or plastically, and the local residual is given by
\begin{equation}
	\label{eq:appendix-Hlocal-matrix-elastic}
	\fH^{\textrm{matrix,elastic}}_n = \mqty[ \feps_{n}^p - \feps_{n-1}^p    \\ 
	p_{n} - p_{n-1}   \\
	\fsigma_n - \fsigma_{n-1} - \C_{\textrm{matrix}} \qty(\feps_{n}-\feps_{n-1} -(\feps_{n}^p-\feps_{n-1}^p) ) \\
	 \lambda_n - \lambda_{n-1}	
	],
\end{equation}
and
\begin{equation}
		\label{eq:appendix-Hlocal-matrix-plastic}
		\fH^{\textrm{matrix,plastic}}_n = \mqty[ \feps_{n}^p - \feps_{n-1}^p - \qty(\lambda_n - \lambda_{n-1})\vb{N}_n  \\ 
		p_{n} - p_{n-1} - \qty(\lambda_n - \lambda_{n-1})\sqrt{\frac{2}{3}} \\
		\fsigma_n - \fsigma_{n-1} - \C_{\textrm{matrix}} \qty(\feps_{n}-\feps_{n-1} -(\feps_{n}^p-\feps_{n-1}^p) ) \\
		\| \fsigma_n^{\textrm{dev}}\| - \sqrt{\frac{2}{3}}\qty(\sigma_y + H p_n)	
		 ],
\end{equation}	
respectively, with the normal $\vb{N}$ to the yield surface computed from the deviatoric stress $\fsigma^{\textrm{dev}}$ via
\begin{equation}
	\label{eq:appendix-normal-flow}
	\vb{N} = \frac{\fsigma^{\textrm{dev}} } { \| \fsigma^{\textrm{dev} }\|}.
\end{equation}

For the matrix, the Jacobian blocks on the elastic branch are
\begin{subequations}
	\label{eq:appendix-Jacobian-Matrix-elastic}
	\begin{align}
		\fH^{\textrm{matrix,elastic},\fv}_n&= \mqty[\vb{I}_{6\times 6} & \vb{0}_{6 \times 1} & \vb{0}_{6 \times 6} & \vb{0}_{6 \times 1} \\
							\vb{0}_{1\times 6} & 1 & \vb{0}_{1\times 6} & 0 \\
							\C_{\textrm{matrix}} & \vb{0}_{6 \times 1} & \vb{I}_{6\times 6} & \vb{0}_{6 \times 1} \\
							\vb{0}_{1 \times 6} & 0 & \vb{0}_{1 \times 6} & 1		
							] , \\
		\fH^{\textrm{matrix,elastic},\feps}_n&= \mqty[ \vb{0}_{6 \times 6} \\
								\vb{0}_{1\times 6} \\
								-\C_{\textrm{matrix}} \\
								\vb{0}_{1\times 6}	],\\
		\fH^{\textrm{matrix,elastic},\fv,-}_n&= \mqty[-\vb{I}_{6\times 6} & \vb{0}_{6 \times 1} & \vb{0}_{6 \times 6} & \vb{0}_{6 \times 1} \\
							\vb{0}_{1\times 6} & -1 & \vb{0}_{1\times 6} & 0 \\
							-\C_{\textrm{matrix}} & \vb{0}_{6 \times 1} & -\vb{I}_{6\times 6} & \vb{0}_{6 \times 1} \\
							\vb{0}_{1 \times 6} & 0 & \vb{0}_{1 \times 6} & -1		
							], \\
		\fH^{\textrm{matrix,elastic},\feps,-}_n&= \mqty[ \vb{0}_{6 \times 6} \\
							\vb{0}_{1\times 6} \\
							\C_{\textrm{matrix}} \\
							\vb{0}_{1\times 6}	].
	\end{align}
\end{subequations}
Meanwhile, on the plastic branch, the Jacobian blocks are obtained as
\begin{subequations}
	\label{eq:appendix-Jacobian-Matrix-plastic}
	\begin{align}
		\fH^{\textrm{matrix,plastic},\fv}_n&= \mqty[\vb{I}_{6\times 6} & \vb{0}_{6 \times 1} & -\qty(\lambda_n - \lambda_{n-1})\frac{\mathbf{P}_{\mathrm{dev}}-\vb{N}_n \otimes \vb{N}_n }{\| \fsigma_{n}^{\textrm{dev}} \|} &  -\vb{N}_n \\
		\vb{0}_{1\times 6} & 1 & \vb{0}_{1\times 6} & -\sqrt{\frac{2}{3}} \\
		\C_{\textrm{matrix}} & \vb{0}_{6 \times 1} & \vb{I}_{6\times 6} & \vb{0}_{6 \times 1} \\
		\vb{0}_{1 \times 6} & -\sqrt{\frac{2}{3}}H & \vb{N}^T_n & 0		
		] , \\
		\fH^{\textrm{matrix,plastic},\feps}_n&= \mqty[ \vb{0}_{6 \times 6} \\
		\vb{0}_{1\times 6} \\
		-\C_{\textrm{matrix}} \\
		\vb{0}_{1\times 6}	],\\
		\fH^{\textrm{matrix,plastic},\fv,-}_n&= \mqty[-\vb{I}_{6\times 6} & \vb{0}_{6 \times 1} & \vb{0}_{6 \times 6} & \vb{N}_n \\
		\vb{0}_{1\times 6} & -1 & \vb{0}_{1\times 6} & \sqrt{\frac{2}{3}} \\
		-\C_{\textrm{matrix}} & \vb{0}_{6 \times 1} & -\vb{I}_{6\times 6} & \vb{0}_{6 \times 1} \\
		\vb{0}_{1 \times 6} & 0 & \vb{0}_{1 \times 6} & 0		
		], \\
		\fH^{\textrm{matrix,plastic},\feps,-}_n&= \mqty[ \vb{0}_{6 \times 6} \\
		\vb{0}_{1\times 6} \\
		\C_{\textrm{matrix}} \\
		\vb{0}_{1\times 6}	],
	\end{align}
\end{subequations}
where $\mathbf{P}_{\mathrm{dev}}$ denotes the matrix representation of the deviatoric projector. The branch is selected by the standard return-mapping algorithm~\cite{wriggers2008nonlinear}. Both the direct and the adjoint differentiation described in Section~\ref{sec:inelastic-training} evaluate the derivatives of the active branch, which are smooth within either the elastic or the plastic regime. We did not encounter evaluation points located exactly on the elastic-plastic boundary, where the derivatives are not defined.

\section{Rank-8 effective-stress trajectories for the unseen loading paths}
\label{app:online_prediction_effective_stress}
The following figures compare the \DMN{e}{8} and \DMN{i}{8} predictions with the FFT reference for all six independent effective-stress components. The component-wise plots complement the global trajectory errors reported in Table~\ref{tab:dmn_ood_trajectory_error}. For a stress component $\alpha$ at the time step $n$, the error is defined by
\begin{equation}
	\label{eq:appendix-componentWise-error}
	e_{\alpha,n} = \frac{\left | \bar{\sigma}_{\alpha,n}^{\textrm{DMN}} -\bar{\sigma}_{\alpha,n}^{\textrm{FFT}} \right | } {\max_{\substack{m=1,\ldots,N_t\\ \beta=1,\ldots,6}} \left | \bar{\sigma}_{\beta,m}^{\textrm{FFT}} \right |+ 10^{-12} }.
\end{equation}

\begin{figure}[p]
	\begin{center}
		\begin{subfigure}{\textwidth}
			\centering
			\includegraphics[height=0.022\textheight]{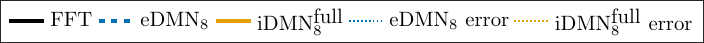}
		\end{subfigure}
	\end{center}
	\centering
	\begin{subfigure}{0.48\textwidth}\includegraphics[width=\linewidth]{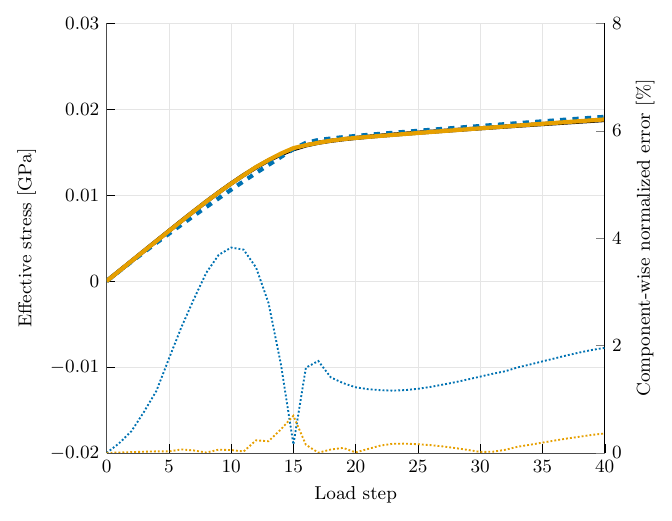}\caption{$\sigma_{11}$}\end{subfigure}\hfill
	\begin{subfigure}{0.48\textwidth}\includegraphics[width=\linewidth]{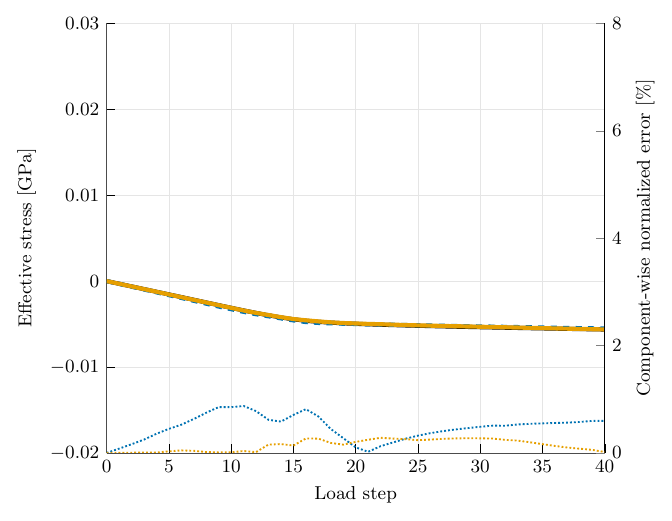}\caption{$\sigma_{22}$}\end{subfigure}
	\begin{subfigure}{0.48\textwidth}\includegraphics[width=\linewidth]{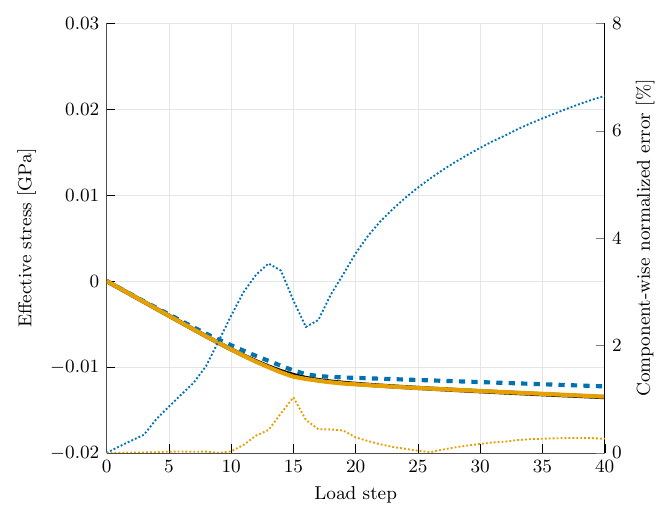}\caption{$\sigma_{33}$}\end{subfigure}\hfill
	\begin{subfigure}{0.48\textwidth}\includegraphics[width=\linewidth]{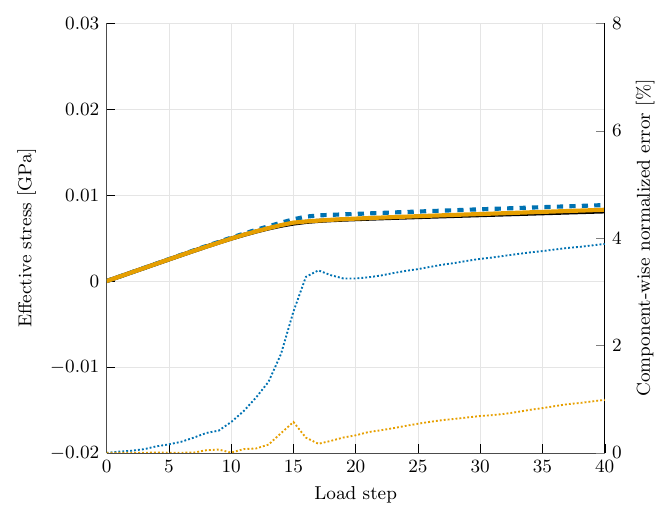}\caption{$\sigma_{23}$}\end{subfigure}
	\begin{subfigure}{0.48\textwidth}\includegraphics[width=\linewidth]{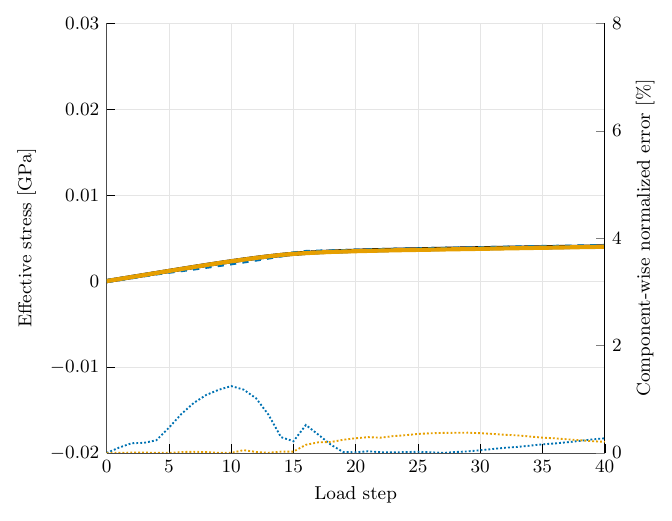}\caption{$\sigma_{13}$}\end{subfigure}\hfill
	\begin{subfigure}{0.48\textwidth}\includegraphics[width=\linewidth]{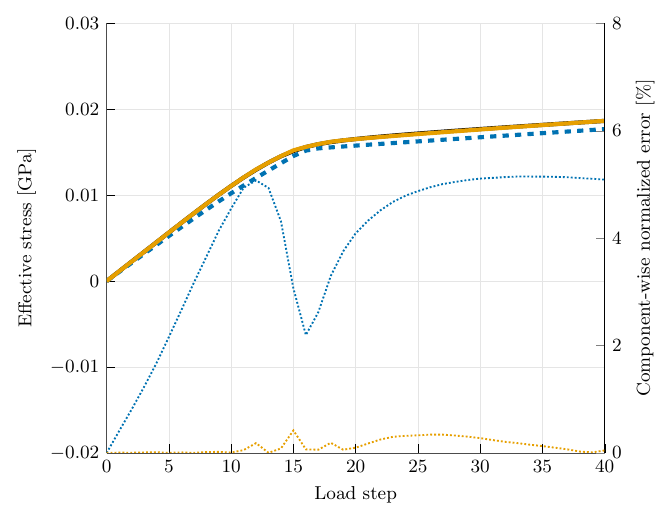}\caption{$\sigma_{12}$}\end{subfigure}
	\caption{Rank-8 effective-stress predictions for the proportional ramp-up OOD path.}
	\label{fig:ood_rampup_rank8}
\end{figure}

\begin{figure}[p]
	\begin{center}
		\begin{subfigure}{\textwidth}
			\centering
			\includegraphics[height=0.022\textheight]{Gen2_results/tex/ood_rank8/legend.pdf}
		\end{subfigure}
	\end{center}
	\centering
	\begin{subfigure}{0.48\textwidth}\includegraphics[width=\linewidth]{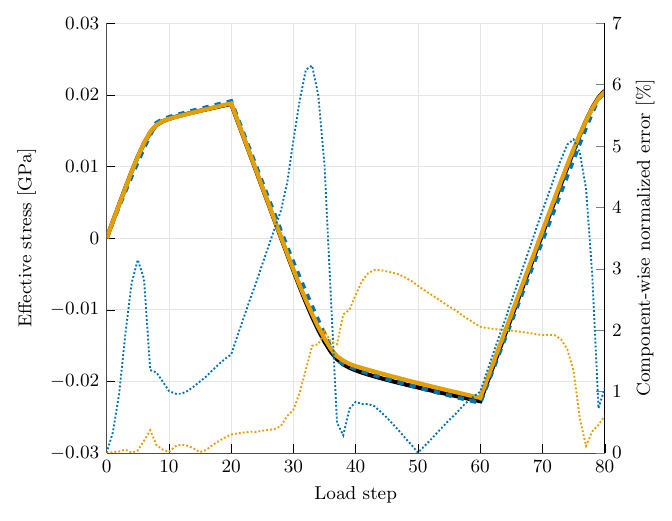}\caption{$\sigma_{11}$}\end{subfigure}\hfill
	\begin{subfigure}{0.48\textwidth}\includegraphics[width=\linewidth]{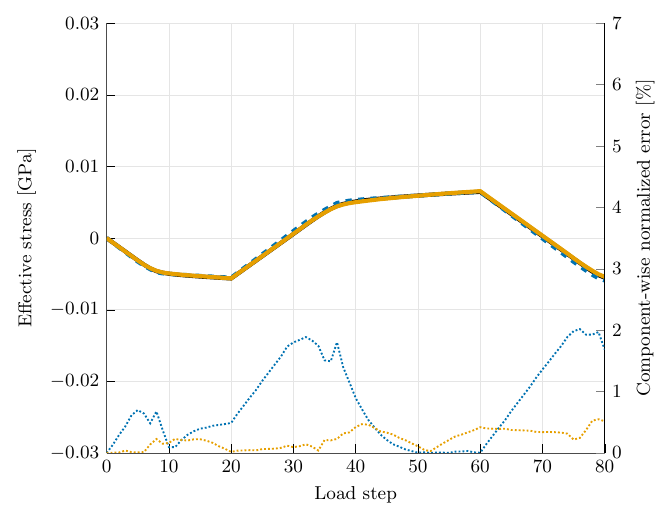}\caption{$\sigma_{22}$}\end{subfigure}
	\begin{subfigure}{0.48\textwidth}\includegraphics[width=\linewidth]{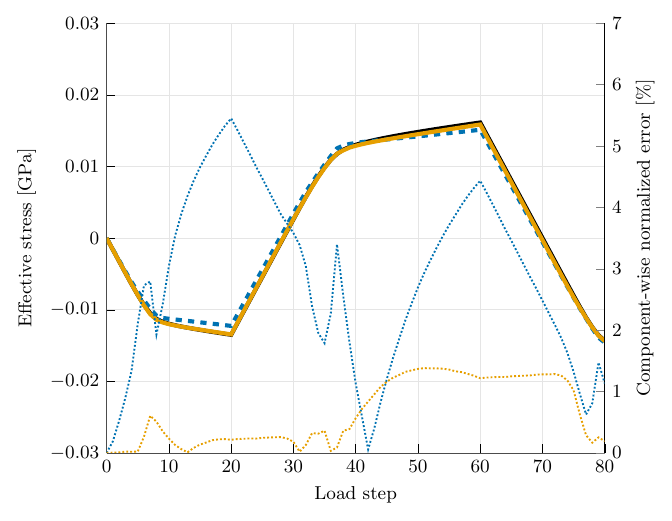}\caption{$\sigma_{33}$}\end{subfigure}\hfill
	\begin{subfigure}{0.48\textwidth}\includegraphics[width=\linewidth]{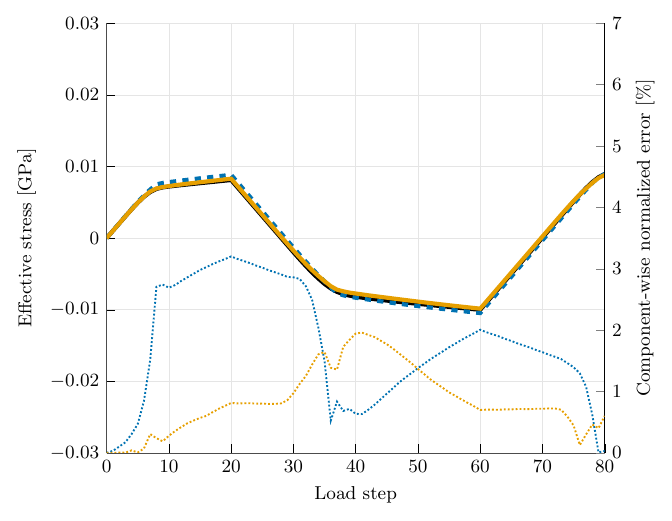}\caption{$\sigma_{23}$}\end{subfigure}
	\begin{subfigure}{0.48\textwidth}\includegraphics[width=\linewidth]{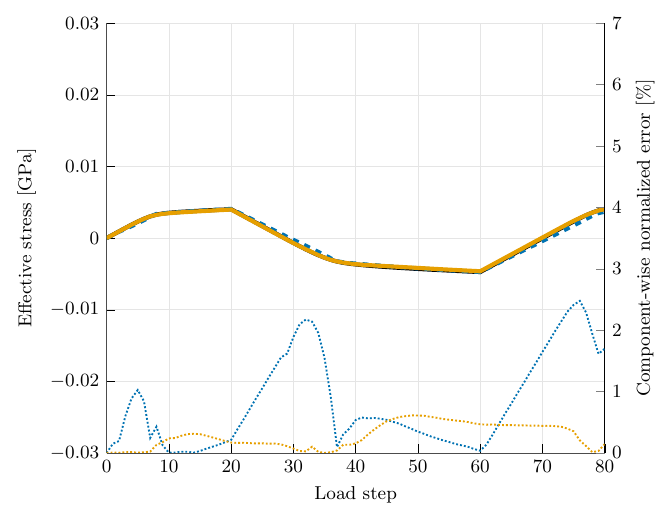}\caption{$\sigma_{13}$}\end{subfigure}\hfill
	\begin{subfigure}{0.48\textwidth}\includegraphics[width=\linewidth]{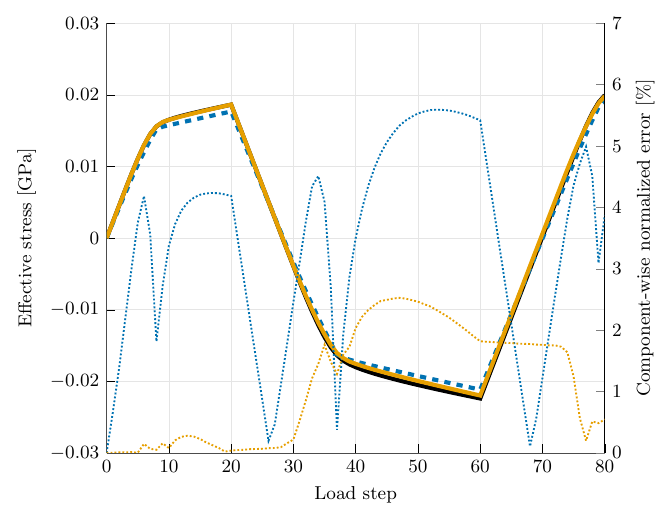}\caption{$\sigma_{12}$}\end{subfigure}
	\caption{Rank-8 effective-stress predictions for the cyclic OOD path.}
	\label{fig:ood_cycle_rank8}
\end{figure}

\begin{figure}[p]
	\begin{center}
		\begin{subfigure}{\textwidth}
			\centering
			\includegraphics[height=0.022\textheight]{Gen2_results/tex/ood_rank8/legend.pdf}
		\end{subfigure}
	\end{center}
	\centering
	\begin{subfigure}{0.48\textwidth}\includegraphics[width=\linewidth]{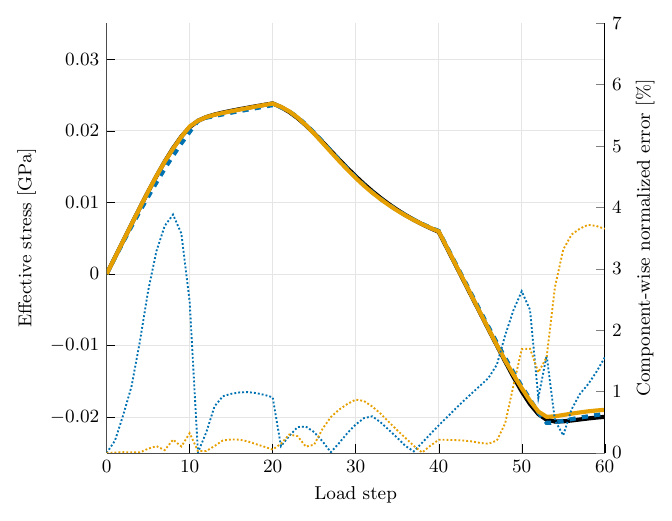}\caption{$\sigma_{11}$}\end{subfigure}\hfill
	\begin{subfigure}{0.48\textwidth}\includegraphics[width=\linewidth]{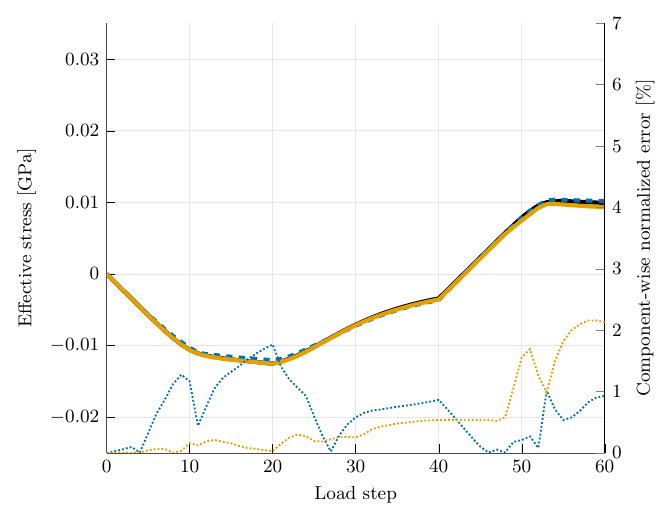}\caption{$\sigma_{22}$}\end{subfigure}
	\begin{subfigure}{0.48\textwidth}\includegraphics[width=\linewidth]{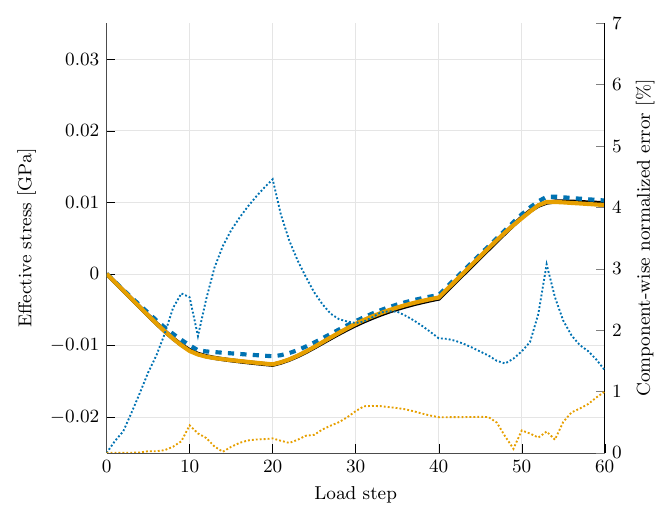}\caption{$\sigma_{33}$}\end{subfigure}\hfill
	\begin{subfigure}{0.48\textwidth}\includegraphics[width=\linewidth]{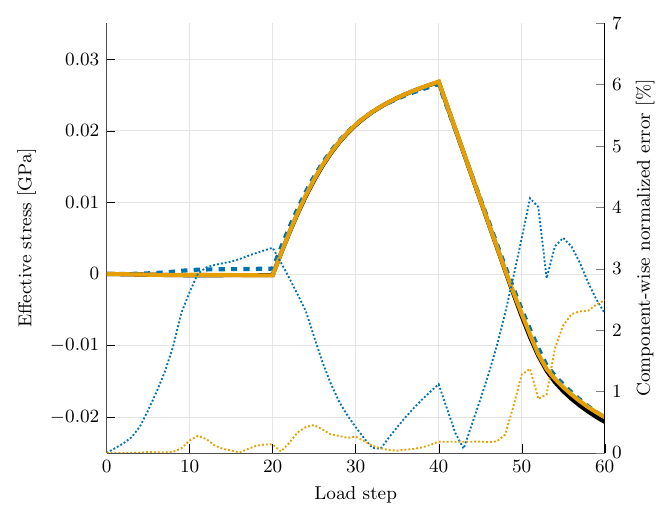}\caption{$\sigma_{23}$}\end{subfigure}
	\begin{subfigure}{0.48\textwidth}\includegraphics[width=\linewidth]{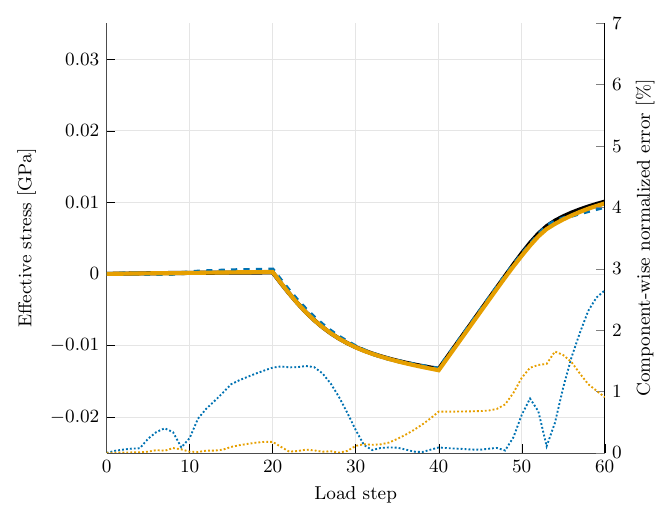}\caption{$\sigma_{13}$}\end{subfigure}\hfill
	\begin{subfigure}{0.48\textwidth}\includegraphics[width=\linewidth]{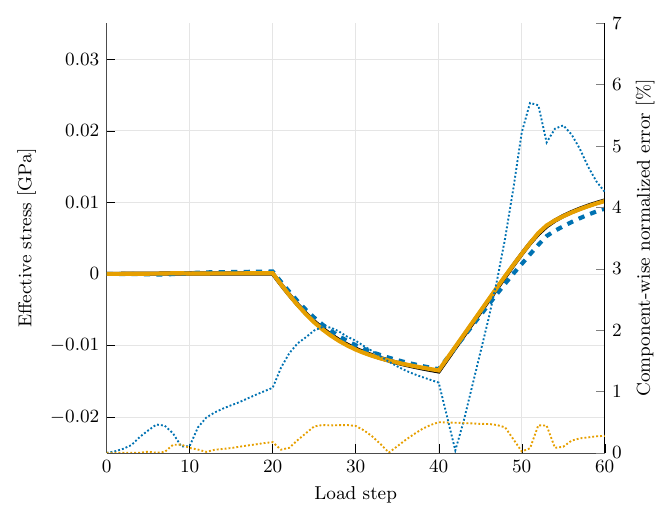}\caption{$\sigma_{12}$}\end{subfigure}
	\caption{Rank-8 effective-stress predictions for the non-proportional OOD path.}
	\label{fig:ood_shear_rank8}
\end{figure}

\section{Material-transfer trajectory errors}
\label{app:material_transfer}
Table~\ref{tab:material_transfer_all} reports the complete set of 225 global trajectory-relative errors used in Section~\ref{sec:material-transfer}. 
\input{Gen2_results/ortho_revision/material_transfer/R5_run/table8_all.tex}

\bibliographystyle{ieeetr}
{\footnotesize \bibliography{literature}}
		
	\end{document}

%% file: commands.tex
\usepackage{bm}
\usepackage[bbgreekl]{mathbbol}

\newcommand{\R}{\mathds{R}}

\newcommand{\C}{\mathds{C}}

\newcommand{\Sym}[1]{\textrm{Sym}(#1)}

\DeclareMathOperator{\dev}{\textrm{dev}}

\newcommand{\mean}[1]{\left\langle {#1} \right\rangle}

\renewcommand{\div}{\textrm{div }}

\newcommand{\sty}[1]{\bm{#1}}

\newcommand{\fa}{\sty{ a}}

\newcommand{\fn}{\sty{ n}}

\newcommand{\fu}{\sty{ u}}
\newcommand{\fv}{\sty{ v}}
\newcommand{\fw}{\sty{ w}}

\newcommand{\fy}{\sty{ y}}

\newcommand{\fE}{\sty{ E}}
\newcommand{\fF}{\sty{ F}}

\newcommand{\fH}{\sty{ H}}
\newcommand{\fI}{\sty{ I}}

\newcommand{\fK}{\sty{ K}}

\newcommand{\fQ}{\bm{ Q}}
\newcommand{\fR}{\sty{ R}}
\newcommand{\fS}{\sty{ S}}

\newcommand{\ftau}{\bm{\tau}}

\newcommand{\feps}{\bm{\varepsilon}}
\newcommand{\fsigma}{\bm{\sigma}}
\newcommand{\fxi}{\bm{\xi}}
\newcommand{\feta}{\bm{\eta}}

\DeclareMathAlphabet{\mathbbold}{U}{bbold}{m}{n}
\newcommand{\ffGamma}{\mathbbold{\Gamma}}

\newcommand{\fgam}{\bm{\gamma}}
\newcommand{\ftheta}{\bm{\theta}}
\DeclareMathOperator{\sym}{sym}

%% file: Gen2_results/ortho_revision/nonlinear_test/tex/iDMN_study/table_iDMN_B_rank_sensitivity_upgraded.tex
\begin{table}[!ht]
    \centering
    \caption{Statistics of the effective-stress sample error $e_s$ on the inelastic test set. All values are given in percent.}
    \label{tab:errors_combined_sigeff_testdata}

    \setlength{\tabcolsep}{3.8pt}
    \renewcommand{\arraystretch}{1.1}

    \begin{tabular}{c|ccc|ccc|ccc|ccc|ccc}
        \toprule
        \multirow{2}{*}{Rank} & \multicolumn{3}{c|}{eDMN} & \multicolumn{3}{c|}{iDMN$^{128}$} & \multicolumn{3}{c|}{iDMN$^{256}$} & \multicolumn{3}{c|}{iDMN$^{512}$} & \multicolumn{3}{c}{iDMN$^{899}$} \\
        \cmidrule(lr){2-4}\cmidrule(lr){5-7}\cmidrule(lr){8-10}\cmidrule(lr){11-13}\cmidrule(lr){14-16}
        & max & mean & std & max & mean & std & max & mean & std & max & mean & std & max & mean & std \\
        \midrule

        4 & 22.44 & 16.28 & 4.33 & 4.06 & 3.75 & 0.30 & 3.98 & 3.68 & 0.27 & 3.79 & 3.60 & 0.18 & 3.97 & 3.64 & 0.33 \\
        5 & 17.12 & 11.66 & 4.23 & 2.42 & 2.12 & 0.31 & 2.21 & 2.02 & 0.20 & 2.28 & 1.99 & 0.30 & 2.12 & 2.00 & 0.14 \\
        6 & 9.94 & 7.50 & 1.59 & 1.53 & 1.36 & 0.15 & 1.31 & 1.25 & 0.06 & 1.30 & 1.26 & 0.03 & 1.27 & 1.18 & 0.08 \\
        7 & 6.55 & 6.01 & 0.53 & 0.89 & 0.85 & 0.04 & 0.81 & 0.78 & 0.02 & 0.75 & 0.73 & 0.02 & 0.75 & 0.74 & 0.02 \\
        8 & 5.82 & 5.12 & 0.43 & 0.63 & 0.61 & 0.02 & 0.56 & 0.55 & 0.02 & 0.52 & 0.50 & 0.02 & 0.52 & 0.50 & 0.02 \\

        \bottomrule
    \end{tabular}
\end{table}

%% file: Gen2_results/ortho_revision/computational_cost/tex/iDMN_study/table_elapsed_seconds_comparison.tex
\begin{table}[!ht]
    \centering
    \caption{Statistics of the training elapsed time $t_{\mathrm{elapsed}}$ across the eDMN and iDMN campaigns. All values are given in seconds.}
    \label{tab:elapsed_seconds_combined}

    \setlength{\tabcolsep}{3.8pt}
    \renewcommand{\arraystretch}{1.1}
	\resizebox{\textwidth}{!}{
    \begin{tabular}{c|ccc|ccc|ccc|ccc|ccc}
        \toprule
        \multirow{2}{*}{Rank} & \multicolumn{3}{c|}{eDMN} & \multicolumn{3}{c|}{iDMN$^{128}$} & \multicolumn{3}{c|}{iDMN$^{256}$} & \multicolumn{3}{c|}{iDMN$^{512}$} & \multicolumn{3}{c}{iDMN$^{899}$} \\
        \cmidrule(lr){2-4}\cmidrule(lr){5-7}\cmidrule(lr){8-10}\cmidrule(lr){11-13}\cmidrule(lr){14-16}
         & max & mean & std & max & mean & std & max & mean & std & max & mean & std & max & mean & std \\
        \midrule

        4 & 77.00 & 76.47 & 0.38 & 576.82 & 484.18 & 107.16 & 635.15 & 540.15 & 126.52 & 673.47 & 511.24 & 147.74 & 564.77 & 438.78 & 109.70 \\
        5 & 83.78 & 82.67 & 0.62 & 1136.57 & 975.95 & 191.57 & 1348.38 & 1332.50 & 14.11 & 1372.98 & 1363.04 & 12.39 & 1369.14 & 1304.77 & 97.72 \\
        6 & 88.76 & 87.97 & 0.45 & 3140.60 & 2948.16 & 301.76 & 3043.29 & 2996.84 & 73.01 & 3136.23 & 2956.20 & 245.93 & 3132.54 & 3112.60 & 28.41 \\
        7 & 95.44 & 94.25 & 1.47 & 8888.53 & 8819.80 & 71.04 & 8817.01 & 8057.47 & 1263.36 & 8993.45 & 8901.72 & 123.21 & 8999.23 & 8617.79 & 470.48 \\
        8 & 106.73 & 102.83 & 3.63 & 35603.67 & 33575.98 & 3431.28 & 35926.38 & 35705.91 & 213.88 & 36092.48 & 35869.18 & 211.37 & 35780.80 & 35708.36 & 64.78 \\

        \bottomrule
    \end{tabular}
}
\end{table}

%% file: Gen2_results/ortho_revision/ood/analysis/table_ood_gajek_all_seeds.tex
\begin{table}[!ht]
    \centering
    \caption{Global trajectory relative $L_2$ error on the three OOD loading paths for the Gajek-trained eDMN and iDMN$^{\text{full}}$ rank sweep. Values are mean $\pm$ sample standard deviation (\%) across five independent eDMN training seeds and three iDMN training seeds per rank.}
    \label{tab:dmn_ood_trajectory_error}
    \begin{tabular}{c|cc|cc|cc}
        \toprule
        \multirow{2}{*}{Rank}
        & \multicolumn{2}{c|}{Ramp-up}
        & \multicolumn{2}{c|}{Cyclic}
        & \multicolumn{2}{c}{Shear} \\
        \cmidrule(lr){2-3}\cmidrule(lr){4-5}\cmidrule(lr){6-7}
        & eDMN & iDMN$^{\text{full}}$ & eDMN & iDMN$^{\text{full}}$ & eDMN & iDMN$^{\text{full}}$ \\
        \midrule
        4 & $15.10 \pm 3.78$ & $4.21 \pm 0.82$ & $17.00 \pm 3.77$ & $5.97 \pm 0.82$ & $16.16 \pm 4.80$ & $6.51 \pm 1.19$ \\
        5 & $10.82 \pm 4.06$ & $1.37 \pm 0.22$ & $13.22 \pm 3.27$ & $3.24 \pm 0.49$ & $14.45 \pm 6.76$ & $4.48 \pm 0.93$ \\
        6 & $8.11 \pm 2.77$ & $1.22 \pm 0.21$ & $10.28 \pm 1.41$ & $2.95 \pm 0.31$ & $8.66 \pm 0.89$ & $3.02 \pm 0.28$ \\
        7 & $4.88 \pm 1.16$ & $0.71 \pm 0.06$ & $6.88 \pm 0.79$ & $2.49 \pm 0.12$ & $7.25 \pm 1.63$ & $2.54 \pm 0.43$ \\
        8 & $5.16 \pm 1.00$ & $0.48 \pm 0.12$ & $6.68 \pm 0.99$ & $2.21 \pm 0.12$ & $5.79 \pm 1.32$ & $1.77 \pm 0.24$ \\
        \bottomrule
    \end{tabular}
\end{table}

%% file: Gen2_results/ortho_revision/material_transfer/R5_run/table7_edmn_idmn128.tex
\begin{table}[!ht]
\centering
\caption{Trajectory-relative effective-stress errors for perturbed matrix properties under cyclic loading. Values are mean $\pm$ sample standard deviation (\%) across independent training seeds: $n=5$ for eDMN and $n=3$ for each iDMN configuration.}
\label{tab:material_transfer_edmn_idmn128}
\setlength{\tabcolsep}{3.2pt}
%\renewcommand{\arraystretch}{1.08}
%\scriptsize
\resizebox{\textwidth}{!}{%
\begin{tabular}{lcrrrrrrrrr}
\toprule
Model & Rank & Nom. & $E_m^{-}$ & $E_m^{+}$ & $\nu_m^{-}$ & $\nu_m^{+}$ & $\sigma_y^{-}$ & $\sigma_y^{+}$ & $H^{-}$ & $H^{+}$ \\
\midrule
eDMN & 4 & $17.00 \pm 3.77$ & $16.54 \pm 3.29$ & $17.54 \pm 4.25$ & $15.85 \pm 2.97$ & $18.50 \pm 4.86$ & $17.76 \pm 4.14$ & $16.39 \pm 3.45$ & $17.17 \pm 3.94$ & $16.86 \pm 3.61$ \\
eDMN & 5 & $13.22 \pm 3.27$ & $13.14 \pm 2.99$ & $13.35 \pm 3.55$ & $12.56 \pm 2.71$ & $14.07 \pm 4.04$ & $13.67 \pm 3.51$ & $12.87 \pm 3.07$ & $13.33 \pm 3.30$ & $13.11 \pm 3.25$ \\
eDMN & 6 & $10.28 \pm 1.41$ & $10.46 \pm 1.37$ & $10.16 \pm 1.45$ & $10.18 \pm 1.45$ & $10.39 \pm 1.38$ & $10.50 \pm 1.36$ & $10.13 \pm 1.43$ & $10.24 \pm 1.44$ & $10.32 \pm 1.39$ \\
eDMN & 7 & $6.88 \pm 0.79$ & $7.05 \pm 0.73$ & $6.79 \pm 0.86$ & $6.81 \pm 0.71$ & $6.99 \pm 0.89$ & $7.05 \pm 0.81$ & $6.78 \pm 0.76$ & $6.95 \pm 0.83$ & $6.83 \pm 0.76$ \\
eDMN & 8 & $6.68 \pm 0.99$ & $6.82 \pm 0.95$ & $6.61 \pm 1.04$ & $6.65 \pm 1.00$ & $6.71 \pm 1.00$ & $6.83 \pm 1.12$ & $6.59 \pm 0.89$ & $6.73 \pm 0.97$ & $6.63 \pm 1.01$ \\
\midrule
iDMN$^{128}$ & 4 & $5.99 \pm 0.69$ & $6.02 \pm 0.49$ & $5.90 \pm 0.77$ & $6.00 \pm 0.65$ & $6.04 \pm 0.77$ & $6.18 \pm 0.85$ & $5.77 \pm 0.46$ & $6.11 \pm 0.67$ & $5.88 \pm 0.70$ \\
iDMN$^{128}$ & 5 & $3.45 \pm 0.18$ & $3.35 \pm 0.14$ & $3.60 \pm 0.21$ & $3.53 \pm 0.21$ & $3.40 \pm 0.16$ & $3.73 \pm 0.21$ & $3.25 \pm 0.17$ & $3.42 \pm 0.17$ & $3.48 \pm 0.19$ \\
iDMN$^{128}$ & 6 & $3.10 \pm 0.11$ & $2.98 \pm 0.09$ & $3.24 \pm 0.13$ & $3.15 \pm 0.07$ & $3.06 \pm 0.18$ & $3.31 \pm 0.09$ & $2.95 \pm 0.13$ & $3.05 \pm 0.14$ & $3.16 \pm 0.08$ \\
iDMN$^{128}$ & 7 & $2.39 \pm 0.14$ & $2.30 \pm 0.14$ & $2.49 \pm 0.13$ & $2.42 \pm 0.09$ & $2.36 \pm 0.19$ & $2.58 \pm 0.13$ & $2.24 \pm 0.14$ & $2.33 \pm 0.10$ & $2.44 \pm 0.17$ \\
iDMN$^{128}$ & 8 & $2.27 \pm 0.10$ & $2.21 \pm 0.09$ & $2.34 \pm 0.10$ & $2.30 \pm 0.12$ & $2.24 \pm 0.09$ & $2.44 \pm 0.10$ & $2.14 \pm 0.09$ & $2.20 \pm 0.10$ & $2.34 \pm 0.10$ \\
\bottomrule
\end{tabular}%
}
\end{table}

%% file: Gen2_results/ortho_revision/material_transfer/R5_run/table8_all.tex
\begin{table}[p]
\centering
\caption{Trajectory-relative effective-stress errors for perturbed matrix properties under cyclic loading. Values are mean $\pm$ sample standard deviation (\%) across independent training seeds: $n=5$ for eDMN and $n=3$ for each iDMN configuration.}
\label{tab:material_transfer_all}
\setlength{\tabcolsep}{2.5pt}
%\renewcommand{\arraystretch}{1.02}
%\scriptsize
\resizebox{\textwidth}{!}{%
\begin{tabular}{lcrrrrrrrrr}
\toprule
Model & Rank & Nom. & $E_m^{-}$ & $E_m^{+}$ & $\nu_m^{-}$ & $\nu_m^{+}$ & $\sigma_y^{-}$ & $\sigma_y^{+}$ & $H^{-}$ & $H^{+}$ \\
\midrule
eDMN & 4 & $17.00 \pm 3.77$ & $16.54 \pm 3.29$ & $17.54 \pm 4.25$ & $15.85 \pm 2.97$ & $18.50 \pm 4.86$ & $17.76 \pm 4.14$ & $16.39 \pm 3.45$ & $17.17 \pm 3.94$ & $16.86 \pm 3.61$ \\
eDMN & 5 & $13.22 \pm 3.27$ & $13.14 \pm 2.99$ & $13.35 \pm 3.55$ & $12.56 \pm 2.71$ & $14.07 \pm 4.04$ & $13.67 \pm 3.51$ & $12.87 \pm 3.07$ & $13.33 \pm 3.30$ & $13.11 \pm 3.25$ \\
eDMN & 6 & $10.28 \pm 1.41$ & $10.46 \pm 1.37$ & $10.16 \pm 1.45$ & $10.18 \pm 1.45$ & $10.39 \pm 1.38$ & $10.50 \pm 1.36$ & $10.13 \pm 1.43$ & $10.24 \pm 1.44$ & $10.32 \pm 1.39$ \\
eDMN & 7 & $6.88 \pm 0.79$ & $7.05 \pm 0.73$ & $6.79 \pm 0.86$ & $6.81 \pm 0.71$ & $6.99 \pm 0.89$ & $7.05 \pm 0.81$ & $6.78 \pm 0.76$ & $6.95 \pm 0.83$ & $6.83 \pm 0.76$ \\
eDMN & 8 & $6.68 \pm 0.99$ & $6.82 \pm 0.95$ & $6.61 \pm 1.04$ & $6.65 \pm 1.00$ & $6.71 \pm 1.00$ & $6.83 \pm 1.12$ & $6.59 \pm 0.89$ & $6.73 \pm 0.97$ & $6.63 \pm 1.01$ \\
\midrule
iDMN$^{128}$ & 4 & $5.99 \pm 0.69$ & $6.02 \pm 0.49$ & $5.90 \pm 0.77$ & $6.00 \pm 0.65$ & $6.04 \pm 0.77$ & $6.18 \pm 0.85$ & $5.77 \pm 0.46$ & $6.11 \pm 0.67$ & $5.88 \pm 0.70$ \\
iDMN$^{128}$ & 5 & $3.45 \pm 0.18$ & $3.35 \pm 0.14$ & $3.60 \pm 0.21$ & $3.53 \pm 0.21$ & $3.40 \pm 0.16$ & $3.73 \pm 0.21$ & $3.25 \pm 0.17$ & $3.42 \pm 0.17$ & $3.48 \pm 0.19$ \\
iDMN$^{128}$ & 6 & $3.10 \pm 0.11$ & $2.98 \pm 0.09$ & $3.24 \pm 0.13$ & $3.15 \pm 0.07$ & $3.06 \pm 0.18$ & $3.31 \pm 0.09$ & $2.95 \pm 0.13$ & $3.05 \pm 0.14$ & $3.16 \pm 0.08$ \\
iDMN$^{128}$ & 7 & $2.39 \pm 0.14$ & $2.30 \pm 0.14$ & $2.49 \pm 0.13$ & $2.42 \pm 0.09$ & $2.36 \pm 0.19$ & $2.58 \pm 0.13$ & $2.24 \pm 0.14$ & $2.33 \pm 0.10$ & $2.44 \pm 0.17$ \\
iDMN$^{128}$ & 8 & $2.27 \pm 0.10$ & $2.21 \pm 0.09$ & $2.34 \pm 0.10$ & $2.30 \pm 0.12$ & $2.24 \pm 0.09$ & $2.44 \pm 0.10$ & $2.14 \pm 0.09$ & $2.20 \pm 0.10$ & $2.34 \pm 0.10$ \\
\midrule
iDMN$^{256}$ & 4 & $5.43 \pm 0.85$ & $5.42 \pm 0.93$ & $5.49 \pm 0.77$ & $5.44 \pm 0.81$ & $5.47 \pm 0.89$ & $5.69 \pm 0.88$ & $5.22 \pm 0.83$ & $5.60 \pm 0.87$ & $5.28 \pm 0.84$ \\
iDMN$^{256}$ & 5 & $3.31 \pm 0.21$ & $3.23 \pm 0.22$ & $3.43 \pm 0.20$ & $3.45 \pm 0.30$ & $3.18 \pm 0.14$ & $3.51 \pm 0.26$ & $3.19 \pm 0.17$ & $3.28 \pm 0.21$ & $3.34 \pm 0.21$ \\
iDMN$^{256}$ & 6 & $2.94 \pm 0.32$ & $2.80 \pm 0.30$ & $3.07 \pm 0.34$ & $2.95 \pm 0.31$ & $2.93 \pm 0.33$ & $3.14 \pm 0.32$ & $2.77 \pm 0.32$ & $2.87 \pm 0.36$ & $3.00 \pm 0.29$ \\
iDMN$^{256}$ & 7 & $2.69 \pm 0.04$ & $2.57 \pm 0.02$ & $2.82 \pm 0.05$ & $2.71 \pm 0.08$ & $2.68 \pm 0.08$ & $2.92 \pm 0.04$ & $2.51 \pm 0.04$ & $2.64 \pm 0.04$ & $2.74 \pm 0.04$ \\
iDMN$^{256}$ & 8 & $2.34 \pm 0.24$ & $2.29 \pm 0.20$ & $2.39 \pm 0.28$ & $2.40 \pm 0.21$ & $2.28 \pm 0.28$ & $2.52 \pm 0.27$ & $2.19 \pm 0.22$ & $2.27 \pm 0.23$ & $2.41 \pm 0.25$ \\
\midrule
iDMN$^{512}$ & 4 & $5.34 \pm 1.52$ & $5.33 \pm 1.55$ & $5.42 \pm 1.48$ & $5.42 \pm 1.40$ & $5.32 \pm 1.63$ & $5.64 \pm 1.63$ & $5.15 \pm 1.40$ & $5.42 \pm 1.57$ & $5.27 \pm 1.47$ \\
iDMN$^{512}$ & 5 & $3.25 \pm 0.57$ & $3.15 \pm 0.58$ & $3.39 \pm 0.58$ & $3.38 \pm 0.62$ & $3.15 \pm 0.53$ & $3.49 \pm 0.58$ & $3.08 \pm 0.59$ & $3.24 \pm 0.54$ & $3.27 \pm 0.61$ \\
iDMN$^{512}$ & 6 & $2.94 \pm 0.07$ & $2.84 \pm 0.12$ & $3.06 \pm 0.04$ & $2.98 \pm 0.12$ & $2.90 \pm 0.03$ & $3.17 \pm 0.08$ & $2.77 \pm 0.08$ & $2.86 \pm 0.08$ & $3.02 \pm 0.08$ \\
iDMN$^{512}$ & 7 & $2.57 \pm 0.09$ & $2.44 \pm 0.07$ & $2.70 \pm 0.10$ & $2.58 \pm 0.14$ & $2.56 \pm 0.05$ & $2.79 \pm 0.11$ & $2.39 \pm 0.06$ & $2.51 \pm 0.09$ & $2.63 \pm 0.09$ \\
iDMN$^{512}$ & 8 & $2.29 \pm 0.09$ & $2.25 \pm 0.11$ & $2.35 \pm 0.07$ & $2.34 \pm 0.06$ & $2.25 \pm 0.12$ & $2.48 \pm 0.08$ & $2.15 \pm 0.09$ & $2.23 \pm 0.09$ & $2.35 \pm 0.09$ \\
\midrule
iDMN$^{\mathrm{full}}$ & 4 & $5.97 \pm 0.82$ & $6.10 \pm 0.71$ & $5.93 \pm 0.93$ & $6.00 \pm 0.76$ & $5.98 \pm 0.90$ & $6.10 \pm 1.10$ & $5.92 \pm 0.59$ & $6.10 \pm 0.86$ & $5.85 \pm 0.78$ \\
iDMN$^{\mathrm{full}}$ & 5 & $3.24 \pm 0.49$ & $3.18 \pm 0.50$ & $3.38 \pm 0.49$ & $3.35 \pm 0.52$ & $3.16 \pm 0.47$ & $3.50 \pm 0.48$ & $3.08 \pm 0.52$ & $3.25 \pm 0.49$ & $3.25 \pm 0.50$ \\
iDMN$^{\mathrm{full}}$ & 6 & $2.95 \pm 0.31$ & $2.87 \pm 0.30$ & $3.06 \pm 0.35$ & $3.02 \pm 0.30$ & $2.89 \pm 0.34$ & $3.18 \pm 0.38$ & $2.78 \pm 0.27$ & $2.90 \pm 0.38$ & $3.01 \pm 0.26$ \\
iDMN$^{\mathrm{full}}$ & 7 & $2.49 \pm 0.12$ & $2.37 \pm 0.12$ & $2.62 \pm 0.12$ & $2.50 \pm 0.15$ & $2.49 \pm 0.10$ & $2.71 \pm 0.15$ & $2.32 \pm 0.09$ & $2.44 \pm 0.13$ & $2.54 \pm 0.12$ \\
iDMN$^{\mathrm{full}}$ & 8 & $2.21 \pm 0.12$ & $2.16 \pm 0.09$ & $2.26 \pm 0.15$ & $2.25 \pm 0.14$ & $2.17 \pm 0.13$ & $2.38 \pm 0.15$ & $2.07 \pm 0.10$ & $2.15 \pm 0.10$ & $2.27 \pm 0.14$ \\
\bottomrule
\end{tabular}%
}
\end{table}